%% file: ms_arXiv.tex
\pdfoutput=1 
\documentclass[]{spie}  %>>> use for US letter paper
\usepackage[]{graphicx}
\usepackage{amssymb, amsmath}

\newcommand{\lya}{Ly$\alpha$}
\newcommand{\ha}{H$\alpha$}
\newcommand{\hb}{H$\beta$}
\newcommand{\oii}{[O II]}
\newcommand{\oiii}{[O III]}
\newcommand{\nii}{[N II]}

\newcommand{\sii}{[S II]}
\newcommand{\arcmin}{$^{\prime}$}
\newcommand{\arcsec}{$^{\prime\prime}$}
\newcommand{\degree}{$^{\circ}$}
\newcommand{\sbcgs}{erg s$^{-1}$ cm$^{-2}$ arcsec$^{-2}$}
\title{LRS2: the new facility low resolution integral field spectrograph for the Hobby-Eberly Telescope\footnote{$\;$The Hobby-Eberly Telescope is operated by McDonald Observatory on behalf of the University of Texas at Austin, the Pennsylvania State University, Ludwig-Maximillians-Universit\"{a}t M\"{u}nchen, and Georg-August-Universit\"{a}t G\"{o}ttingen.}} 

\author{Taylor S. Chonis\supit{a}, Gary J. Hill\supit{a,b}, Hanshin Lee\supit{b}, Sarah E. Tuttle\supit{b}, Brian L. Vattiat\supit{b}
\skiplinehalf
\supit{a}The University of Texas at Austin, Department of Astronomy, 2515 Speedway, Stop C1400, Austin, TX, USA 78712; \\
\supit{b}The University of Texas at Austin, McDonald Observatory, 2515 Speedway, Stop C1402, Austin, TX, USA 78712; \\
}

\authorinfo{Further author information: (Send correspondence to T.S.C.)\\T.S.C.: E-mail: tschonis@astro.as.utexas.edu, Telephone: 1-512-471-1495}

\begin{document} 
  \maketitle 
%%%%%%%%%%%%%%%%%%%%%%%%%%%%%%%%%%%%%%%%%%%%%%%%%%%%%%%%%%%%% 

\begin{abstract}
The second generation Low Resolution Spectrograph (LRS2) is a new facility instrument for the Hobby-Eberly Telescope (HET). Based on the design of the Visible Integral-field Replicable Unit Spectrograph (VIRUS), which is the new flagship instrument for carrying out the HET Dark Energy Experiment (HETDEX), LRS2 provides integral field spectroscopy for a seeing-limited field of 12\arcsec$\times$6\arcsec. For LRS2, the replicable design of VIRUS has been leveraged to gain broad wavelength coverage from 370 nm to 1.0 $\mu$m, spread between two fiber-fed dual-channel spectrographs, each of which can operate as an independent instrument. The blue spectrograph, LRS2-B, covers $370 \leq \lambda\; \mathrm{(nm)} \leq 470$ and $460 \leq \lambda\; \mathrm{(nm)} \leq 700$ at fixed resolving powers of $R = \lambda / \delta\lambda \approx 1900$ and 1100, respectively, while the red spectrograph, LRS2-R, covers $650 \leq \lambda\; \mathrm{(nm)} \leq 842$ and $818 \leq \lambda\; \mathrm{(nm)} \leq 1050$ with both of its channels having $R\approx1800$. In this paper, we present a detailed description of the instrument's design in which we focus on the departures from the basic VIRUS framework. The primary modifications include the fore-optics that are used to feed the fiber integral field units at unity fill-factor, the cameras' correcting optics and detectors, and the volume phase holographic grisms. We also present a model of the instrument's sensitivity and a description of specific science cases that have driven the design of LRS2, including systematically studying the spatially resolved properties of extended \lya\ blobs at $2<z<3$. LRS2 will provide a powerful spectroscopic follow-up platform for large surveys such as HETDEX. 
\end{abstract}

%>>>> Include a list of keywords after the abstract 
%%%%%%%%%%%%%%%%%%%%%%%%%%%%%%%%%%%%%%%%%%%%%%%%%%%%%%%%%%%%%

%%%%%%%%%%%%%%%%%%%%%%%%%%%%%%%%%%%%%%%%%%%%%%%%%%%%%%%%%%%%%
\section{INTRODUCTION} \label{sec:intro}  % \label{} allows reference (\ref{}) to this section
Broadband low resolution spectroscopy is a powerful general purpose tool employed at nearly all major optical astronomical observatories to serve the needs of a wide range of research programs. For the past decade, the 9.2 m Hobby-Eberly Telescope (HET\cite{Ramsey98}) has provided this capability to its users through the Marcario Low Resolution Spectrograph (LRS; Ref. \citenum{Hill98}), which has been a workhorse facility instrument that has provided the majority of the telescope's citations over that time period. The HET is a unique telescope and employs an 11 m hexagonal-shaped, segmented spherical primary mirror fixed at a zenith angle of 35\degree. The telescope structure can rotate 360\degree\ in azimuth, allowing the telescope to access $\sim70$\% of the sky visible at McDonald Observatory. To track astronomical objects, a robotic $x$-$y$ tracker sweeps the pupil across the over-sized primary mirror rather than moving the whole telescope structure. The Prime Focus Instrument Package (PFIP) that rides on the tracker contains the spherical aberration correcting optics, metrology instrumentation\cite{Lee12a}, and fiber feeds for the science instruments. Currently, the HET is undergoing a major wide-field upgrade (WFU; Ref. \citenum{Hill14b}) in preparation for the upcoming HET Dark Energy eXperiment (HETDEX; Ref. \citenum{Hill08a}). The WFU, which includes a new, more advanced tracker\cite{Good14a} and PFIP\cite{Vattiat14} assembly, is centered around a new wide-field spherical aberration corrector (WFC\cite{Oh14}) that will increase the telescope's field of view to 22\arcmin, increase the pupil size to 10 m, feature improved throughput, metrology, and image quality, and optimize the telescope to efficiently feed fiber-coupled instruments at a faster focal ratio ($f$/3.65). Currently, the LRS rides on the tracker as a part of the PFIP and is fed directly by the HET spherical aberration corrector. However, due to the physical and optical constraints imposed by the new PFIP assembly and WFC, respectively, the current LRS instrument is incompatible with the WFU and must either be replaced or modified to be fiber-coupled to the focal plane of the upgraded telescope. 

The upcoming HETDEX survey will amass a sample of $\sim$0.8 million \lya\ emitting galaxies (LAEs) to be used as tracers of large-scale structure for constraining dark energy and measuring its possible evolution from $1.9 < z < 3.5$. To carry out the 120 night blind spectroscopic survey covering a 420 degree$^{2}$ field (9 Gpc$^{3}$), the HET will be outfitted with a revolutionary new multiplexed instrument called the Visible Integral Field Replicable Unit Spectrograph (VIRUS; Ref. \citenum{Hill14a}). VIRUS consists of at least 150 copies of a simple fiber-fed integral field spectrograph and is the first optical astronomical instrument to leverage the economies of scale associated with large-scale replication to significantly reduce overall costs. By taking advantage of the large engineering investment already made in VIRUS and its mass production lines\cite{Tuttle14,Marshall14}, a new facility instrument with improved sensitivity can be procured to replace LRS at only a slightly higher cost than modifying the current instrument to be compatible with the WFU. The VIRUS unit spectrograph design is versatile and was intended to be adaptable to a range of spectral resolutions and wavelength coverage configurations. As a result, the VIRUS multiplex advantage can be used to gain broad spectral coverage, which provides an ideal platform on which to build the second generation LRS instrument (hereafter, referred to as LRS2).

The original LRS2 design concept\cite{Lee10} was fed by a 12\arcsec$\times$7\arcsec\ integral field unit (IFU) with unity spatial fill-factor and simultaneously covered $350 < \lambda\:(\mathrm{nm}) < 1100$ at a fixed resolving power of $R = \lambda / \delta\lambda \approx 1800$. The wide spectral coverage is made possible by utilizing two VIRUS unit pairs (i.e., four total spectrograph channels). While much of the design concept discussed in Ref. \citenum{Lee10} remains unchanged in the final LRS2 design presented here, there have been some significant adjustments. As outlined in Ref. \citenum{Chonis12a}, the largest change is that the quad-channel instrument that would have provided simultaneous coverage from $350 < \lambda\: (\mathrm{nm}) < 1100$ has been broken up into two, independent double-spectrographs that will observe the wavelength ranges of $370 < \lambda\: (\mathrm{nm}) < 700$ and $650 < \lambda\: (\mathrm{nm}) < 1050$, respectively. Other than a new, highly customized IFU, the blue optimized spectrograph pair (LRS2-B) requires only modest adaptation of the VIRUS design (the most significant of which is custom dispersing elements) while the red optimized pair (LRS2-R) requires similar modifications to LRS2-B in addition to different optical coatings, optimized refractive camera correcting optics, and custom red-sensitive CCD detectors. LRS2-B and LRS2-R will be built on the VIRUS production line using many components that are common with VIRUS, which will greatly reduce the delivery time and final cost for such a capable instrument. 

As simple, robust general purpose instruments with rapid setup times and high efficiency, LRS2-B and LRS2-R are designed to be highly complementary to VIRUS and to exploit the queue-scheduled\cite{Shetrone07} HET for efficient survey follow-up, synoptic observations, and observations of targets of opportunity (TOOs). In the following section, we outline the scientific utility of the LRS2 spectrographs with an emphasis on the drivers for setting the spectral resolution and wavelength coverage of the four individual spectrograph channels. In $\S$\ref{sec:design}, we provide a technical description of the instrument and focus on the departures from the base design of VIRUS. The operation of the instrument is discussed in $\S$\ref{sec:operation}, and a model of the instrumental sensitivity is presented in $\S$\ref{sec:sensitivity}. Finally, we summarize and provide an outlook towards the completion and commissioning of the instrument in $\S$\ref{sec:status}.
%%%%%%%%%%%%%%%%%%%%%%%%%%%%%%%%%%%%%%%%%%%%%%%%%%%%%%%%%%%%%

%%%%%%%%%%%%%%%%%%%%%%%%%%%%%%%%%%%%%%%%%%%%%%%%%%%%%%%%%%%%%
\section{SCIENTIFIC DRIVERS}\label{sec:science}
LRS2 will enable the continuation of current LRS science programs as well as provide new and improved capabilities. While a VIRUS-based design will not replicate the imaging or multi-object capabilities of LRS, the majority of LRS usage has historically been spectroscopic slit observations of individual small objects ($<10$\arcsec\ in size). Additionally, the imaging capability previously provided by LRS will be assumed by the new acquisition camera\cite{Vattiat14} that is included as part of the new PFIP while the wide field, multi-object spectroscopy will be carried out with VIRUS\cite{Hill14a}. The strongest requirements for LRS2 are broadband spectral coverage with high efficiency (particularly in the red) and excellent sky subtraction. LRS has historically been the dominant instrument in dark time and has been used for the largest variety of science applications among the three current HET instruments\cite{Hill04}. We expect LRS2 to have similar broad application and to increase the efficiency of many of the tasks at which its predecessor excelled. Additionally, the change from a slit to an IFU and the enhanced red sensitivity will enable a range of new science applications. 

LRS2 now consists of two optimized double-spectrographs, each with a sharp dichroic beam splitter separating the light into two optimized spectrograph channels. The possibility of utilizing a third dichroic for simultaneous spectral coverage from the near-UV to the near-IR was considered\cite{Lee10}, but the design was complicated by the tight space constraints on the HET focal plane and the required alignment tolerance. Given the requirements on the desired spectral resolution and the constraints imposed by using VIRUS as a building block for LRS2, introducing a third dichroic cross-over would also complicate the data quality near the \ha\ line. Additionally, the large difference in background levels between the near-UV and the near-IR and the typical spectral energy distributions of common targets makes splitting the total wavelength range with significant overlap around \ha\ a good compromise between science goals and engineering complexity.

\subsection{Specific Science Cases for LRS2-B and LRS2-R}\label{subsec:scicasesi}
Below, we list several high-level descriptions of science cases that LRS2 will contribute to:

\textbf{Targets of Opportunity -} When coupled with the responsive queue scheduling\cite{Shetrone07} of the HET, LRS has proven to be an ideal instrument for time critical studies of transient phenomena such as follow-up spectroscopy of $\gamma$-ray bursts (GRBs) and supernovae. Similar future observations will greatly benefit from the increased spectral coverage of LRS2 and its IFU, which will significantly decrease setup times. This increases observing efficiency since coordinates of transients often arrive with arcsecond-level uncertainties. The IFU will also improve the two-dimensional background subtraction of the underlying host galaxies of GRBs and supernovae. A prime example of LRS TOO observations that LRS2 can perform more efficiently is the completion of the Sloan Digital Sky Survey (SDSS)-II Supernova Survey, for which LRS provided spectra for the sample's most distant objects to $z\sim0.5$, mostly within 24 hours of discovery\cite{Zheng08}. These types of observations help provide cosmological constraints on dark energy at intermediate redshifts\cite{Kessler09}. LRS also has obtained rapid response spectra of GRBs\cite{Schaefer03}, including the brightest ever observed\cite{Racusin08}. GRBs are now being discovered up to $z\sim8$ and provide deep lines of sight through the high redshift intergalactic medium that can be used to probe reionization. The broad spectral coverage, rapid acquisition capability, and high red efficiency of LRS2 will make it ideal for these studies.

\textbf{Synoptic Observations -} Synoptic observations, such as the reverberation mapping\cite{Kaspi07} of active galactic nuclei (AGN) have been well suited to the capabilities of LRS and will continue to be with LRS2. Additionally, LRS2 will efficiently provide extended synoptic observations of supernovae to monitor their spectra as they fade. These types of observations can probe the inner structure of the explosions and provide insights into the explosion physics\cite{Quimby07}. 

\textbf{Single Object Studies and Survey Follow-Up -} With broad wavelength coverage, excellent sky subtraction, and efficient operation in survey mode through the HET queue schedule\cite{Shetrone07}, LRS2 will play a uniquely important role as a follow-up instrument for interesting objects found through upcoming large surveys that will discover an unprecedented number of objects that will require spectroscopic follow-up (e.g., the Dark Energy Survey, Pan-STARRS, and the Large Synoptic Survey Telescope). LRS2 will enable the following examples:
\begin{itemize} \itemsep1pt \parskip0pt \parsep0pt
	\item \textbf{Supernovae:} Thousands of moderate-redshift supernovae will be found in fields that are accessible to the HET. The ability to rapidly obtain spectra of these supernovae will lead to improved redshift determinations, more accurate classification, and reduced systematic errors in the derived Hubble diagrams. Ref. \citenum{Zheng08} shows the power of flexibly-scheduled spectroscopy of supernovae (including many spectra from LRS). As an example, the Dark Energy Survey\cite{DES} will discover and measure light curves for $\gtrsim3000$ Type Ia supernovae in 5 years for $0.3 < z < 1.0$. LRS2 could be used for follow-up spectroscopy of a large number of Type Ia candidates as well as their host galaxies to confirm redshifts. The predicted sensitivity of LRS2 (see $\S$\ref{sec:sensitivity}) is also well matched to the expected supernovae discovery brightness.
	\item \textbf{Brown Dwarfs:} The coldest, low-mass brown dwarfs represent valuable laboratories for studying cool, planet-like atmospheres and provide constraints on the low end of the stellar initial mass function. Spectroscopy from 600 nm to 1 $\mu$m is critical for identifying and studying brown dwarfs since the M and L spectral types are identified by molecular absorption bands in the red\cite{Kirkpatrick99}, \ha\ can be used as a diagnostic of accretion from circumstellar disks\cite{Muzerolle05}, and gravity sensitive absorption lines can place constraints on masses\cite{Cruz07}. A use of LRS in this field has been for the confirmation of brown dwarf candidates identified in photometric surveys of young stellar objects in star-forming regions\cite{Luhman09}. With the increased sensitivity of LRS2-R over the current LRS at the relevant wavelengths, surveys for these young stellar objects with the HET could reach mass limits as low as $\sim5$ $M_{\mathrm{Jupiter}}$.
	\item \textbf{Quasars:} LRS has made important contributions to identifying quasar candidates from current surveys (e.g., Ref. \citenum{Schneider00}). With near-UV to near-IR sensitivity, LRS2 will be superb for spectroscopic follow up of a wide range of imaging surveys from X-ray to infrared and radio wavelengths (e.g. Ref. \citenum{Brand03}). A powerful feature of LRS2 is that the spectral resolution is fixed and the IFU adapts the data to any image quality without losing light at the entrance aperture. Upcoming surveys will yield significant numbers of $z\sim6$-7 quasar candidates brighter than 21 AB mag at 950 nm which is easily within LRS2's reach for confirming and calibrating photometric redshift measurements. 
	\item \textbf{Galaxies and Clusters:} LRS has also contributed to studies of galaxy structure and dynamics\cite{Corsini08}. LRS2 will also do so and will continue to enable emission line diagnostics in the local universe for the standard optical transitions (e.g., \oii, \oiii, \ha, \hb, \nii) while providing two dimensions of spatial information. Upcoming large imaging surveys will also discover many galaxy clusters over a moderate range of redshifts that will be used for cluster-cosmology science and studies of galaxy evolution. LRS2 could be used to survey candidate clusters and at a minimum could provide validation of the cluster-finding algorithms and assess the completeness of galaxy catalogs.
	\item \textbf{HETDEX:} In addition to detecting $\sim1$ million LAEs from $1.9 < z < 3.5$, HETDEX\cite{Hill08a} will also discover as many \oii\ emitters at $z<0.5$. It will also deliver spectra of large numbers of AGN, stars, galaxy clusters, and more down to the sensitivity limit of the SDSS imaging survey. This unique dataset will require follow-up with LRS2 to establish, for example, the classification of emission line objects and extremely metal poor stars, and to study emission line diagnostics over a wider wavelength range. For example, \ha, \nii, and \sii\ emission line diagnostics will be within the LRS2-R bandpass for all HETDEX \oii\ emitters. Understanding galactic winds is also an upcoming field, and HETDEX will provide a substantial sample of blue compact dwarf galaxies at $z<0.5$ where LRS2 can search for the emission line diagnostics of outflows over a wide wavelength at moderate resolving power.
	\item \textbf{\lya\ and High Redshift Galaxies:} With its broad wavelength coverage, LRS2 provides the ability to detect \lya\ emission from $2 < z < 7.6$. The radiative transfer of \lya\ has long been an intense field of study since \lya\ emission is a prominent selection method for star-forming galaxies and quasars in the early universe, it provides a mechanism for confirming redshifts, and can probe reionization\cite{Stark10}. \lya\ radiative transfer is complicated due to the resonant nature of the transition, which effectively scatters the photons through neutral hydrogen in a geometric and frequency diffusion process. This makes the transition very sensitive to gas velocity fields, line of sight neutral gas density, and the presence of dust\cite{Verhamme06,Barnes11}. LRS2 will be useful for follow-up observations of the most interesting LAEs discovered by HETDEX (e.g., largest mass, highest star formation rate, etc.), especially near the epoch when the star formation rate of the universe was near its peak and when most galaxy assembly occurred (i.e., $2 \lesssim z \lesssim 2.7$). This redshift range also places the galaxies' rest-frame optical nebular transitions (e.g., \hb, \oiii, \ha, \nii) into the near-IR atmospheric windows where infrared spectrographs (e.g., MOSFIRE\cite{McLean12}, KMOS\cite{Sharples12}) can provide measurements of systemic redshifts and standard rest-frame optical emission line diagnostics\cite{Finkelstein11}. In conjunction with spectrally resolved \lya\ spectra, these data provide a path towards learning about the physics of these young star-forming galaxies, especially the effects of feedback (e.g., star burst driven winds) from extreme star formation on the galactic environment. This is especially true for the spatially extended \lya\ blobs with sizes as large as 100 kpc (e.g., Ref. \citenum{Yang10}). These objects can be studied in great detail with the LRS2 IFU, and an unprecedented sample of them will be compiled from VIRUS HETDEX observations. We outline this specific science case as an application of our LRS2 sensitivity calculations in $\S$\ref{subsec:sensapplication}.
\end{itemize}

\subsection{Spectrograph Channel Wavelength Coverage}\label{subsec:wavecoverage}
As mentioned, the two LRS2 double-spectrographs have wavelength coverages of $370 < \lambda\: (\mathrm{nm}) < 700$ (LRS2-B) and $650 < \lambda\: (\mathrm{nm}) < 1050$ (LRS2-R), which significantly overlaps around \ha. Due to the fixed VIRUS spectrograph design into which LRS2 is being adapted (see $\S$\ref{subsec:VIRUS}), there is a trade-off between spectral coverage and spectral resolution for each of the four spectrograph channels. Additionally, the cross-over wavelengths of the two dichroic beam splitters must be chosen to minimize the science impact since there is an inevitable loss of signal to noise ratio (SNR) around the wavelengths of the split.

For LRS2-B, there are two primary criteria that set the wavelength range of the dichroic cross-over. First, the LRS ``G2'' configuration\cite{Hill98,Hill03}, which simultaneously observes \hb\ to \ha\ with contingency for a small range in redshift, must be replicated in a single channel. Second, sufficient spectral resolution and wavelength coverage for enabling LAE follow-up studies from $2 \lesssim z \lesssim 2.7$ must be achieved. Additionally, accommodating the broad line widths often observed in AGN is desired. For our purposes, these considerations place the channel transition from $460-470$ nm. Note that the lower limit to the spectrograph's coverage contains \oii\ at $z=0$, which extends the current capability of LRS toward the UV by $\sim50$ nm. We thus call the bluer LRS2-B channel the ``UV Arm''. The redder ``G2'' replicating channel is referred to as the ``Orange Arm''.

For LRS2-R, the specific choice of the transition wavelength is less important than maintaining an adequately high spectral resolution in both channels to ensure good sky subtraction in the far red where the strong night sky emission bands dominate the background. It is also desirable to maintain approximately the same spectral resolution between the two channels to allow diagnostic spectral features to be analyzed consistently over the whole wavelength range. These considerations loosely place the channel transition at $\sim850$ nm. We refer to the bluer LRS2-R channel as the ``Red Arm'' and the redder channel as the ``Far Red Arm''.
%%%%%%%%%%%%%%%%%%%%%%%%%%%%%%%%%%%%%%%%%%%%%%%%%%%%%%%%%%%%%

%-------------
   \begin{figure}[t]
   \begin{center}
   \begin{tabular}{c}
   \includegraphics[width=0.98\textwidth]{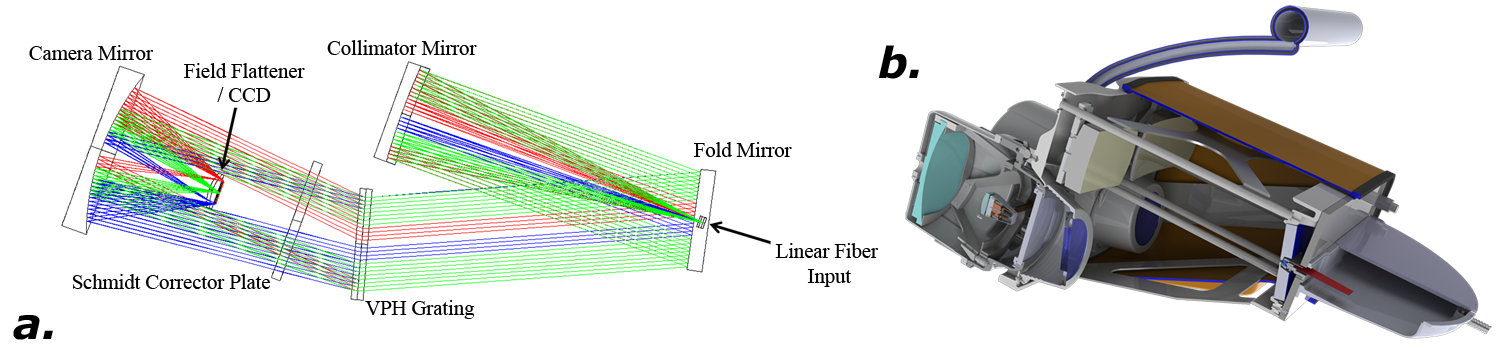}
   \end{tabular}
   \end{center}
   \caption[example] 
   { \label{fig:VIRUS} 
\textit{a}) The optical layout of VIRUS with major components labeled. This basic optical layout is the same for all four LRS2 channels except that the VPH gratings are exchanged for custom VPH grisms in addition to small changes to the aspheric profile prescriptions of the Field Flatteners and/or the Schmidt Corrector Plates. For scale, the collimated beam size is 125 mm. The fiber ``slit'' is oriented out of the page. \textit{b}) A section view of the mechanical design of the VIRUS unit spectrograph pair. The plumbing extending above the spectrograph is part of the liquid nitrogen distribution manifold\cite{Chonis10}. The IFU output head containing the fiber ``slit'' is mounted to the bulkhead at right, and a schematic section of the fibers can be seen in the housing in red. 
}
   \end{figure} 
%------------- 

%%%%%%%%%%%%%%%%%%%%%%%%%%%%%%%%%%%%%%%%%%%%%%%%%%%%%%%%%%%%%
\section{INSTRUMENT DESIGN}\label{sec:design}
LRS2 is based on the VIRUS unit spectrograph design that has been carefully optimized for mass production and benefits greatly from the investment made in its development and production-line engineering. We begin by discussing the key design features of VIRUS that form the basis of the LRS2 spectrographs (see Ref. \citenum{Hill14a} for a complete description). The departures from the VIRUS baseline that morph it into the LRS2 spectrographs are discussed individually in the subsections that follow.

\subsection{VIRUS: The Building Block of LRS2}\label{subsec:VIRUS}
The VIRUS unit spectrograph is designed to be fed by fiber optics and is optimized to accept a $f$/3.65 input beam (it can accept a beam as fast as $f$/3.4 to account for slight misalignments and a small amount of focal ratio degradation in the fibers\cite{Murphy08,Murphy12}). The optical design (see Fig. \ref{fig:VIRUS}$a$) is a double-Schmidt, with the collimator reversed. The Schmidt collimator and camera share a corrector plate and utilize a volume phase holographic (VPH) grating\cite{Chonis14} for dispersion at the pupil. The focal reduction factor from the collimator to the camera is $2.8\times$. The wavelength coverage for HETDEX is $350 < \lambda\:(\mathrm{nm}) < 550$, but the design is pan-chromatic since the optical power is mainly in the two spherical mirrors. For adaptation to different wavelength ranges while maintaining the superb image quality of the base design, the mirrors and transmissive optics can be supplied with different coatings and custom aspheric correcting optics can be fabricated (see $\S$\ref{subsec:collimator} and $\S$\ref{subsec:camera}). The dispersion and wavelength coverage can be further adapted by using grisms instead of a standard VPH grating (see $\S$\ref{subsec:grisms}). VIRUS is designed to be fed by a ``densepak'' IFU, where the fibers are arrayed in a hexagonal pattern with a 1/3 spatial fill-factor at the input from the telescope focal plane. For input into the instrument, the fibers are arranged in a linear ``slit'' pattern (see $\S$\ref{subsec:IFU}). The VIRUS IFU is fed directly at $f$/3.65 by the HET WFC. An observation with VIRUS consists of three exposures, each in one of three dither positions separated by $\sim1$\arcsec\ that fill in the sky coverage gaps between the fibers. The dithers are achieved by precisely moving the focal plane assembly on which the IFUs are mounted rather than moving the telescope itself (see Ref. \citenum{Vattiat14}). The dithering is thus independent of guiding and guarantees precise knowledge of the relative positions of the three exposures that complete the observation. In production, each VIRUS mechanical unit contains a pair of spectrographs in a common housing that are fed by a single IFU (see Fig. \ref{fig:VIRUS}$b$). Since this configuration allows a pair of spectrographs to share common components (such as CCD electronics, vacuum vessels, cryogenic components, etc.), costs are reduced and more efficient use of the limited space available at the telescope structure is allowed for. Two of these unit pairs will form the basis of LRS2. A single prototype VIRUS spectrograph (the Mitchell Spectrograph, formerly VIRUS-P\cite{Hill08b}), has been in regular use as a facility instrument on the McDonald Observatory 2.7 m telescope since 2007. In addition to verifying the VIRUS concept, it has resulted in the successful completion of the HETDEX Pilot Survey (HPS\cite{Adams11}) in preparation for HETDEX. The Mitchell Spectrograph has demonstrated the high throughput, excellent image quality, low ghosting, and exceptional stability of the VIRUS design.

%-------------
   \begin{figure}[t]
   \begin{center}
   \begin{tabular}{c}
   \includegraphics[width=0.9\textwidth]{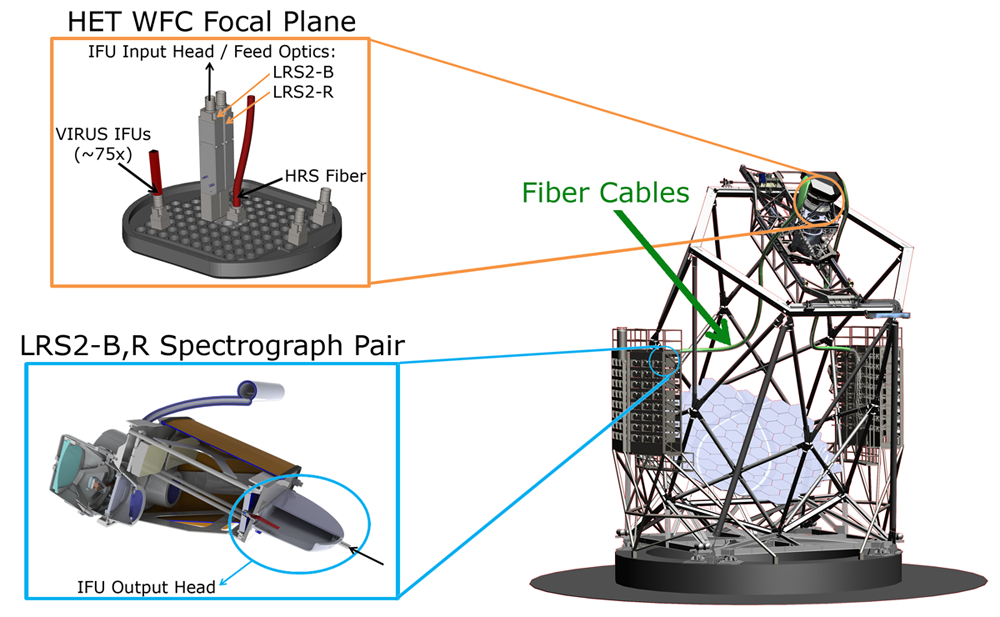}
   \end{tabular}
   \end{center}
   \caption[example] 
   { \label{fig:FiberCable1} 
A schematic of the LRS2 fiber light path. The HET WFC focal plane is located at the top of the telescope inside the PFIP assembly (indicated by the orange circle). The arrangement of VIRUS IFUs, fiber feeds for other HET facility instruments, and the LRS2 IFUs on the fiber feed mounting plate at the WFC focal plane is shown in the orange inset. Only three VIRUS IFUs are shown along with the fiber feed for the High Resolution Spectrograph (HRS) and the IFU feeds for the two LRS2 spectrograph pairs. A VIRUS IFU will be installed in each of the open ports on the mounting plate, which highlights the limited space in which the LRS2 feed optics must be deployed. At the other end of the fiber cable is a spectrograph pair (shown in the blue inset) where the fibers are arranged into two linear ``slits'' within the IFU output head to feed each spectrograph channel. Since the LRS2 spectrographs will be located within the enclosures of the VIRUS Support Structure (see the blue circle on the large box mounted onto the side of the HET structure), the LRS2 fiber cables will be bundled together with the cables from the VIRUS IFUs.
}
   \end{figure} 
%------------- 

\subsection{Feed Optics and Fiber Integral Field Units}\label{subsec:IFU}
The most significant departure from VIRUS for LRS2 is the fiber IFU design and how it is fed by the HET. For VIRUS, the bare 266 $\mu$m (1.5\arcsec) core diameter fibers of the $1/3$ spatial fill-factor IFUs are fed directly by the HET WFC. For LRS2, however, we desire IFUs with $\sim100$\% spatial fill-factor and a sufficient field size to ease target acquisition and provide adequate sky sampling for small objects in all image quality conditions. Additionally, we desire small enough spatial elements to subsample the typical HET image quality of 1.2\arcsec\ full width at half maximum (FWHM), which will require reimaging the telescope's fast $f$/3.65 beam.

\subsubsection{Fiber Cables}\label{subsubsec:IFUcable}
In addition to the two LRS2 IFU feed assemblies (one for LRS2-B, and one for LRS2-R), the focal plane of the upgraded HET is populated by the fiber feeds for the other HET facility instruments\cite{Hill04,Mahadevan14} and at least 75 IFUs that feed the VIRUS spectrograph array\cite{Hill14a} (see Fig. \ref{fig:FiberCable1}). As a result, HET focal plane ``real-estate'' is extremely limited and the strict space restrictions will drive the design of the reimaging optics for feeding the LRS2 IFUs.  

%-------------
   \begin{figure}[t]
   \begin{center}
   \begin{tabular}{c}
   \includegraphics[width=0.9\textwidth]{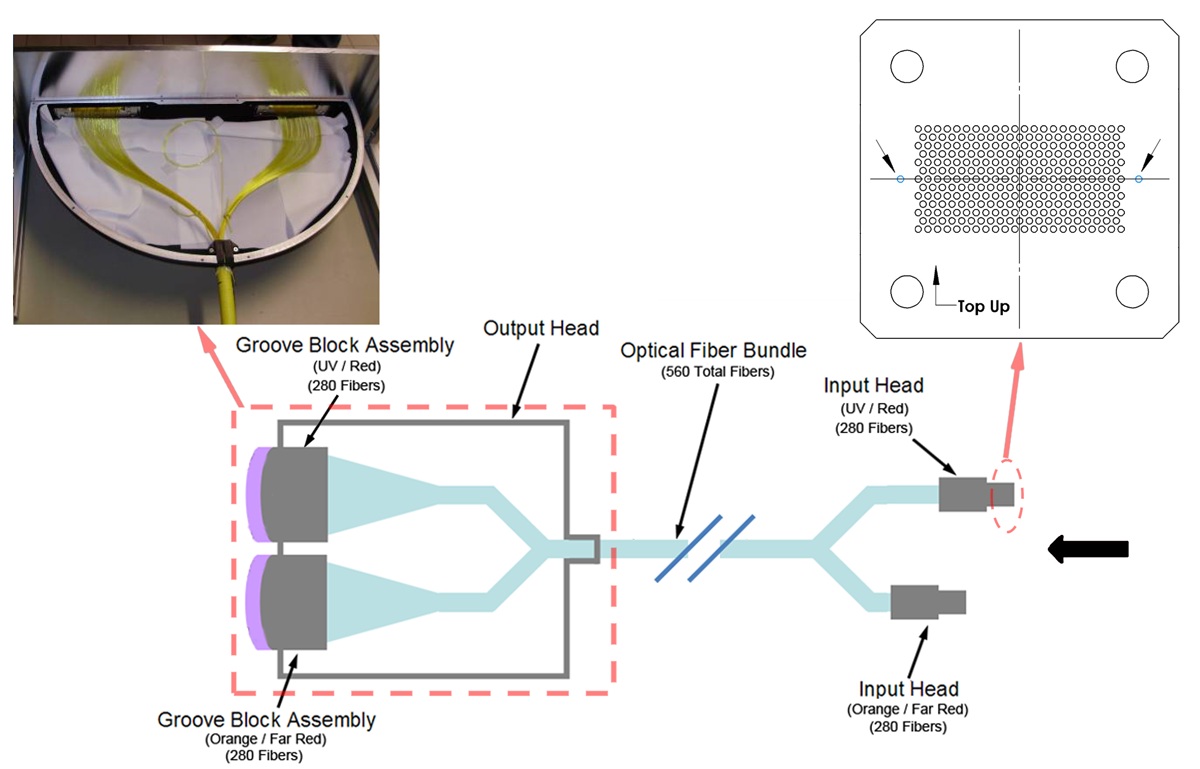}
   \end{tabular}
   \end{center}
   \caption[example] 
   { \label{fig:FiberCable2} 
A schematic of the LRS2 IFU fiber cable layout with major components labeled, as viewed from above. The two fiber cables (one for each LRS2-B and LRS2-R spectrograph pair) each contain two input heads, one for each wavelength-specific spectrograph channel, where the fibers are arrayed into the rectangular pattern with 13 rows alternating between 22 and 21 fibers each. Looking at the input heads along the thick black arrow at right, the fiber layout of the precision drilled hole block is seen detailed at the top right of the figure. The two additional arrowed holes outside of the rectangular fiber arrangement are used as an alignment reference, as discussed in $\S$\ref{subsubsec:IFUmech}. The fibers from the two input heads are bundled together and routed through a single conduit into the output head, where the fibers are again separated and laid out into two linear ``slits'' to feed each spectrograph channel. A photo of the fiber separation and arrangement into the ``slits'' inside of the output head can be seen at the top left of the figure. 
}
   \end{figure} 
%------------- 

Each LRS2 spectrograph pair will rely on the well-established VIRUS IFU cable design that is described in detail in Refs. \citenum{Hill14a} and \citenum{Kelz14}, with the only differences being the fiber core diameter, the number of fibers, and the design of the fiber input head. The VIRUS fiber cable design has been shown to be quite robust, and has successfully passed the equivalent of 10 years of operational moves without displaying any significant decay in its optical properties\cite{Murphy12}. To help achieve higher spectral resolution and the desired spatial sampling, the LRS2 IFUs are constructed with fibers that have 170 $\mu$m core diameter (36\% smaller than VIRUS). The number of spatial elements in the final IFU design is limited by the number of spectra that can be packed onto the CCDs with sufficient separation for clean extraction. For our purposes, this separation must be such that $<30$\% of the peak intensity of the fiber spatial profile is reached at the overlap point. For the VIRUS spectrograph design with 170 $\mu$m core diameter fiber, this number is 280 per spectrograph channel. The overall length of the fiber cable is 18 m. 

For VIRUS, a single input IFU head containing 448 fibers feeds the pair of spectrographs contained within a single VIRUS unit. At the output head of the fiber cable, the 448 fibers are split evenly between the two linear ``slits'' that feed each spectrograph (see Fig. \ref{fig:FiberCable2}). For each LRS2 spectrograph pair, however, the light will be split spectrally using a dichroic before the fibers are initially fed (see below), so each spectrograph channel has its own input head containing 280 fibers. The layout of the 280 fibers on the LRS2 input heads can be seen in Fig. \ref{fig:FiberCable2}, and the mechanical design is discussed below in $\S$\ref{subsubsec:IFUmech}. The fibers from each input head are bundled together through a single conduit and routed to the spectrograph pair. Using the same method as VIRUS at the output head, the fibers are separated again and arranged into the linear ``slits'' to feed the respective wavelength optimized spectrograph channels. This is depicted schematically in Fig. \ref{fig:FiberCable2}. For the LRS2 IFU output head, the hardware assembly is identical to VIRUS except for the precision machined stainless ``groove block'' that sets the angle and spacing between the fibers within the ``slit''.

%-------------
   \begin{figure}[t]
   \begin{center}
   \begin{tabular}{c}
   \includegraphics[width=0.98\textwidth]{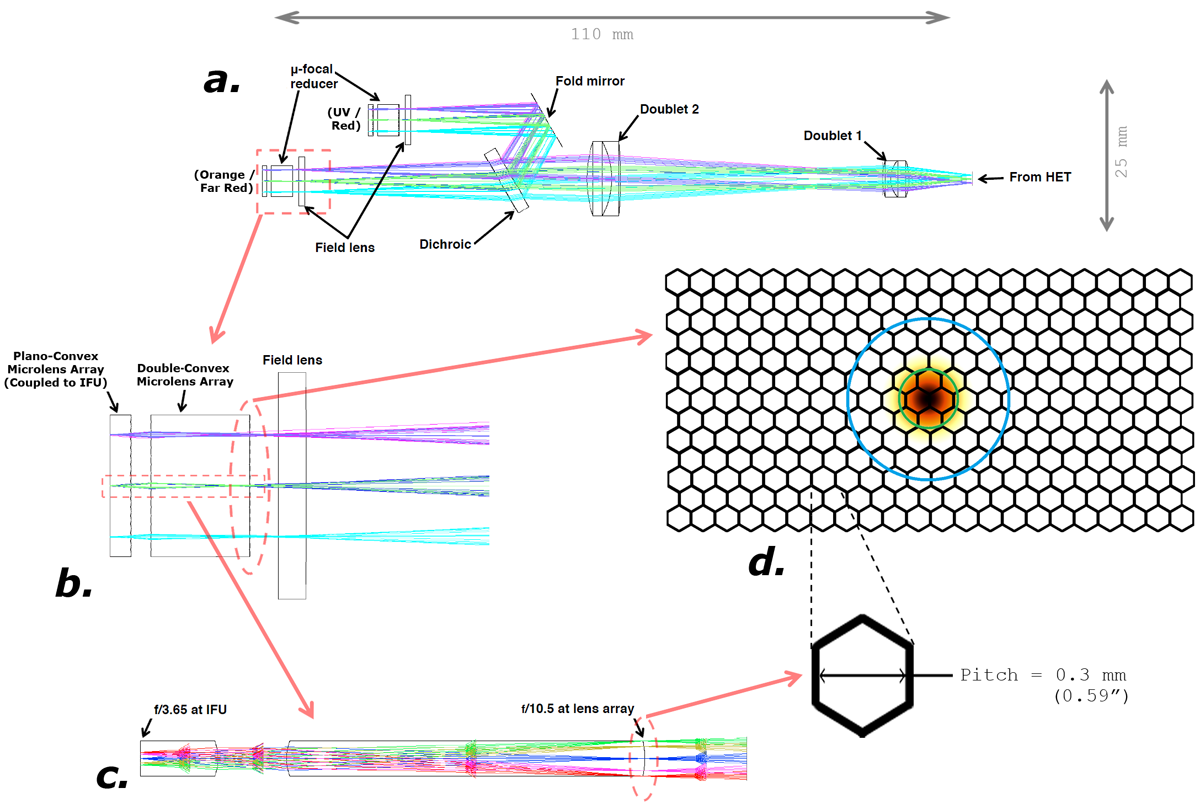}
   \end{tabular}
   \end{center}
   \caption[example] 
   { \label{fig:IFUFeed} 
\textit{a}) The optical design and ray-trace of the LRS2 IFU input optical system. The design is identical for both LRS2-B and LRS2-R, except for the coatings on the lenses, fold mirror, and microlens arrays, and the dichroic design. \textit{b}) Detailed view of the microfocal reducer optics, consisting of a field lens and the two lenslet arrays that provide the field slicing. \textit{c}) The optical path for a single spatial element of the lenslet arrays, showing the conversion of the $f$/10.5 input beam from the doublet relay back to $f$/3.65 to optimally feed the optical fibers. \textit{d}) The layout of the 280 spatial elements of the microlens arrays. Each spatial element is hexagonal with a pitch of 0.3 mm (0.59\arcsec). For scale, a 1.5\arcsec\ FWHM Moffat point spread function is shown along with the VIRUS fiber size (1.5\arcsec, shown in green) and the Mitchell Spectrograph fiber size (4.2\arcsec, shown in blue).
}
   \end{figure} 
%------------- 

\subsubsection{Optical Design}\label{subsubsec:IFUoptics}
\begin{figure}[t]
   \begin{center}
   \begin{tabular}{c}
   \includegraphics[width=0.95\textwidth]{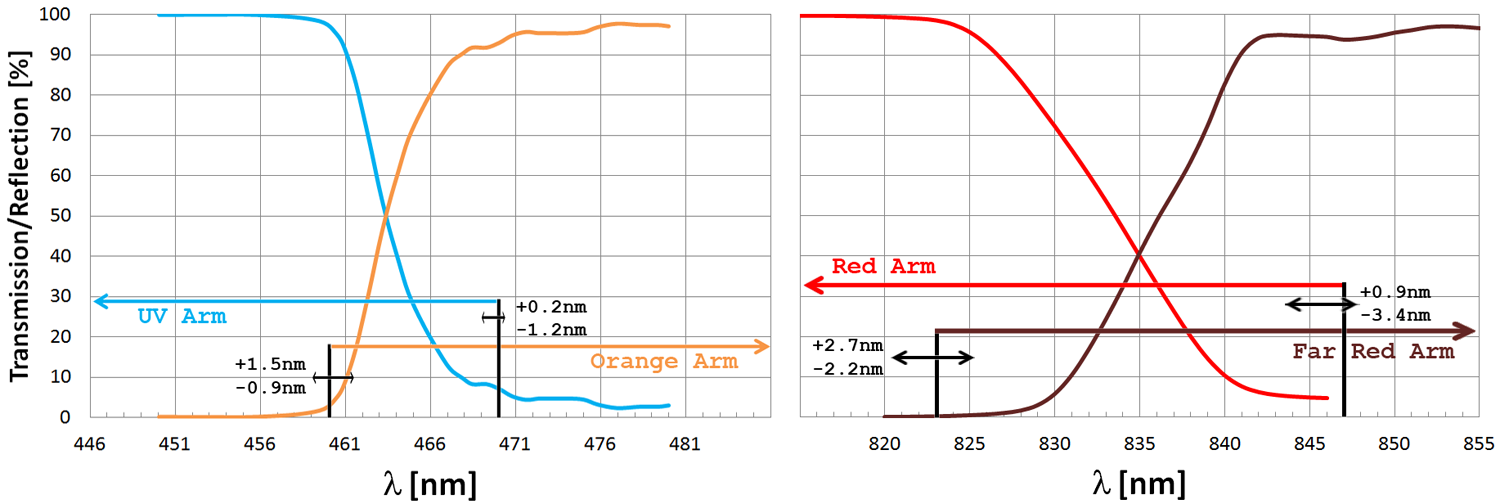}
   \end{tabular}
   \end{center}
   \caption[example] 
   { \label{fig:dichroic} 
The transmissivity and reflectivity of the LRS2-B (left) and LRS2-R (right) dichroic. For each dichroic, the bluer wavelength channel is reflected. The total wavelength range shown in each plot is 40 nm. The designed wavelength cutoff for each spectrograph channel around the transition region is shown in the plots along with the tolerance on the cutoff due to the expected level of precision in the fabrication and assembly of the VPH grisms (see $\S$\ref{subsec:grisms}).
}
   \end{figure}

Fig. \ref{fig:IFUFeed} shows the optical layout of the IFU input optical system for LRS2 (note the scale indicating its small size). The optical design is the same for the LRS2-B and LRS2-R fiber feed optical systems except for the coatings on the lenses, fold mirror, and dichroic. The IFU input optical system consists of a focal extender made with two custom doublets (fabricated from fused silica and LLF1) mounted after the telescope focus to convert the $f$/3.65 telescope beam to $f$/10.5 for finer spatial sampling. The two doublets are 6 and 12 mm in diameter, respectively. The field of view of the focal extender is $\sim14$\arcsec\ in diameter. The image quality of the dual-doublet relay is excellent with 90\% of the diffraction encircled energy enclosed within $<0.25$\arcsec\ diameter across the central 12\arcsec\ of its field of view. An optimized dichroic after the reimaging doublet (i.e., Doublet 2 in Fig. \ref{fig:IFUFeed}) splits the beam by reflecting bluer light and transmitting redder light. For LRS2-B and LRS2-R, the as-built dichroic transition wavelength (i.e., the wavelength at which the transmissivity and reflectivity of the dichroic are equal) is 463.9 nm and 835.0 nm, respectively, as motivated by the science objectives discussed in $\S$\ref{sec:science}. The fold mirror in the bluer arm of each IFU is coated to optimize for the wavelength coverage of that channel. The dichroic is used at a shallow 30\degree\ angle of incidence to help sharpen the transition between the transmission and reflection regions. In Fig. \ref{fig:dichroic}, we show the transmissivity and reflectivity of the LRS2-B and LRS2-R dichroics. For both dichroics at the operating angle of incidence, the transition from reflection to transmission with increasing wavelength is very sharp and decreases from 80\% to 20\% reflectance in 4 nm and 7 nm for LRS2-B and LRS2-R, respectively. 

%-------------
%------------- 

After the dichroic split and subsequent additional reflection for the bluer channel, each channel's beam is focused onto a microfocal reducer, which reimages the $f$/10.5 beam back to $f$/3.65 to directly image the sliced field onto the individual fiber ends, by way of a 8 mm diameter fused silica field lens. As shown by the ray-trace in Fig. \ref{fig:IFUFeed}, the microfocal reducer consists of two fused-silica microlens arrays, which provide the field slicing with a high spatial fill factor of $\gtrsim95$\% and have hexagonal elements with a pitch of 0.3 mm (see Fig. \ref{fig:IFUFeed} for the pitch definition). The pitch size corresponds to 0.59\arcsec\ on the sky, which allows sufficient sampling of the typical seeing at the HET. The IFU layout consists of 13 rows of fibers which alternate between containing 22 or 21 individual fibers. This layout covers a 12.4\arcsec$\times$6.1\arcsec\ field as measured from the centers of the microlens elements (physically, the active area of the microlens array is 6.3 mm $\times$ 3.1 mm). For our purposes, the elongated field is advantageous over a circular microlens arrangement since it provides the opportunity to place potential sky fibers farther from the observed object, depending on its size, shape, and placement on the IFU.  

The first microlens array in the microfocal reducer assembly has lenslets that are double-convex while the second has lenslets that are plano-convex (all convex surfaces are spherical). Since the microimages formed by the individual microfocal reducer lenslets are 120 $\mu$m in size, the optical design allows $\pm25$ $\mu$m of leeway for the fiber IFU-to-microfocal reducer alignment to minimize unnecessary losses. Feeding fibers with direct imaging using the microfocal reducer method described here is more complicated than traditional lenslet-coupled IFUs, which typically form micropupil images on the fiber ends (e.g., Ref. \citenum{AllingtonSmith02}). However, high quality modern optical fibers exhibit little enough focal ratio degradation when fed at $f$/3.65 that the telescope's central obstruction is maintained in the spectrograph for directly illuminated fibers\cite{Murphy08}. Since central obscurations also exist in the LRS2 spectrograph due to its folded collimator and Schmidt camera, the retention of the telescope's central obstruction within the spectrograph results in lower light losses. Another advantage of the microfocal reducer is that it allows the retention of telecentricity through the entire IFU input optical system, which results in less sensitivity to alignment errors that could result in additional losses and effective focal ratio degradation. Modeling of the microfocal reducer-based IFU feed design predicts an overall fiber coupling efficiency $\sim90$\%, including diffraction effects. Table \ref{tab:IFUProp} summarizes the properties of the fiber feed optics and IFU as described in this section. 

\input{t1.tex}

%-------------
   \begin{figure}[t]
   \begin{center}
   \begin{tabular}{c}
   \includegraphics[width=0.9\textwidth]{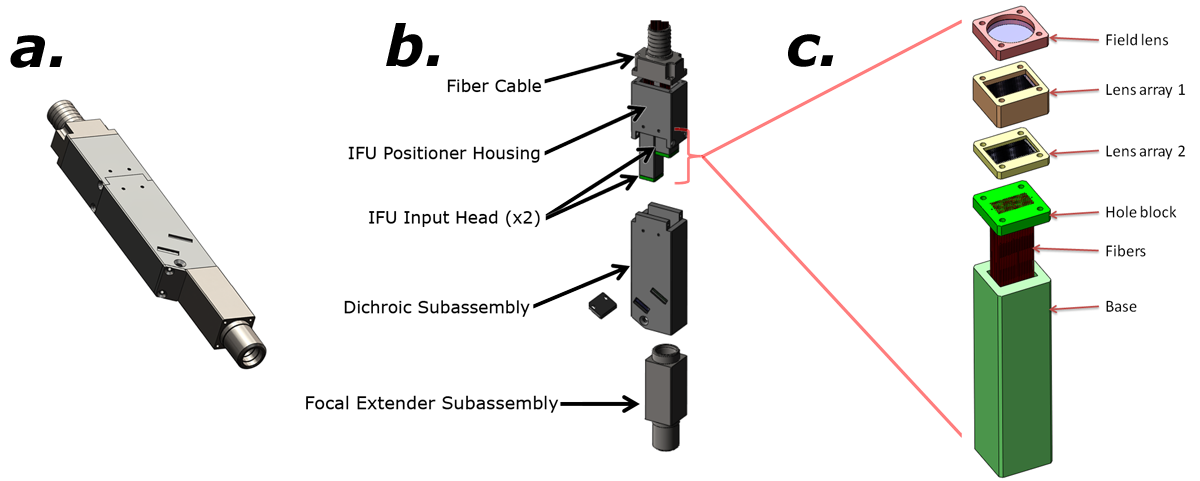}
   \end{tabular}
   \end{center}
   \caption[example] 
   { \label{fig:IFUmech1} 
\textit{a}) The LRS2 IFU input feed mechanical assembly. \textit{b}) An exploded view of the IFU input feed assembly with major subassemblies labeled. \textit{c}) An exploded view of the IFU input head, including the optomechanical components that make up the microfocal reducer.
}
   \end{figure} 
%------------- 

\subsubsection{Mechanical Design and Alignment}\label{subsubsec:IFUmech}
The mechanical design of the IFU input head and feed assembly is shown in Fig. \ref{fig:IFUmech1}. It consists of five subassemblies: the first houses the dual doublet focal extender, the second houses the dichroic and fold mirror, and the third positions the two IFU input heads. The remaining two subassemblies are the two IFU input heads, which include the mechanical framework for positioning the optics of the microfocal reducer.

\begin{figure}[t]
   \begin{center}
   \begin{tabular}{c}
   \includegraphics[width=0.98\textwidth]{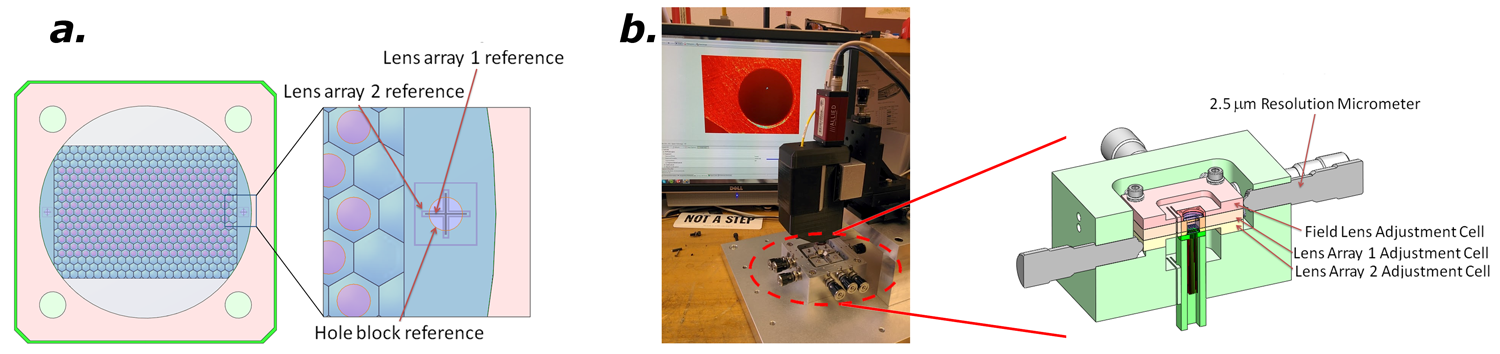}
   \end{tabular}
   \end{center}
   \caption[example] 
   { \label{fig:IFUmech2} 
Alignment of the IFU input head and microfocal reducer optics. \textit{a}) A schematic view of the alignment features used to position the two lenslet arrays with respect to the fiber positions, which are shown as purple circles behind the microlens arrays. In the zoomed-in inset, one of the alignment reference holes on the IFU hole block is shown with the etched reference features on the two microlens arrays. This view is similar to what would be seen by the machine vision camera on the alignment jig. \textit{b}) A photo of the IFU input head alignment jig. The zoomed-in inset shows a section view of the jig's fine-adjustment mechanisms.  
}
   \end{figure}

For the focal extender and dichroic subassemblies, the optics can be positioned and fastened to sufficient accuracy by using mechanical positioning features fabricated with precision machine tolerance. Positioning the optics of the microfocal reducer with sufficient accuracy with respect to the fiber array is much more challenging, however. Fig. \ref{fig:IFUmech1}$c$ shows the components of the IFU input head and microfocal reducer. The 280 fibers are coupled to the alloy hole block in the same way as VIRUS as described in Ref. \citenum{Hill14a}. The hole block is then fastened to the input head base. The field lens and each of the two microlens arrays are epoxied into mounting cells. To align the two microlens arrays and field lens with respect to the positions of the fibers on the hole block, we have developed a custom 8-degree of freedom alignment jig to make fine adjustments of the two microlens arrays and the field lens. The jig is equipped with a machine vision camera that monitors the alignment of reference features drilled into the hole block and precision etched into each of the microlens arrays. The alignment jig and a schematic of the alignment features can be seen in Fig. \ref{fig:IFUmech2}. Once aligned, epoxy is injected into the four large holes in the IFU hole block and the mounting cells for the lenslet arrays and field lens. These can be seen in the exploded view of the IFU input head assembly in Fig. \ref{fig:IFUmech1}$c$.

%-------------
%------------- 

Once the microfocal reducer optics have been aligned to the fiber array, each of the two assembled input heads are then inserted into the IFU position housing. The registration of the two IFUs must be accurate to within $\leq1/10$ of the lenslet pitch for spatially resolved studies of extended objects for which the broad wavelength coverage of the combination of the two spectrograph channels is desired. Such registration accuracy is achieved using precision machine tolerances on two perpendicular inner walls of the IFU position housing and the outside walls of the two input head bases. Fasteners accessible from the outside of the IFU position housing allow the input head bases to be securely positioned against the two reference surfaces. 

\subsection{VPH Grisms}\label{subsec:grisms}
Since the two LRS2 spectrograph pairs are housed inside the mechanical framework of a VIRUS unit, the collimator and camera angles are not free parameters. To adapt such a fixed setup to a wavelength coverage configuration and spectral resolution that differs from the base VIRUS design for a fixed fiber core diameter, we must immerse a standard VPH grating between prisms (i.e., a grism) to tune the beam deviation for various grating fringe densities. For all four LRS2 grisms, the prisms and VPH grating substrates are fabricated from BK7, and the grisms are to be used in first order. Note that the fixed spectrograph design naturally creates a trade-off between the spectral coverage and spectral resolution for a fixed fiber core diameter (e.g., if one desires higher spectral resolution, the channel's wavelength coverage must be reduced). Therefore, the designs of the VPH grisms are determined by the desired wavelength coverage for each spectrograph channel. Considering the delivered sharpness of the dichroic transitions (see Fig. \ref{fig:dichroic}), we have elected to leave 10 and 24 nm of overlap between the two channels of the LRS2-B and LRS2-R spectrograph pairs, respectively. Given the resulting wavelength coverage for each channel and the fixed VIRUS detector size used for LRS2, the linear dispersion of the grism is known and a good estimate of the required fringe density can be determined.

%-------------
   \begin{figure}[t]
   \begin{center}
   \begin{tabular}{c}
   \includegraphics[width=0.9\textwidth]{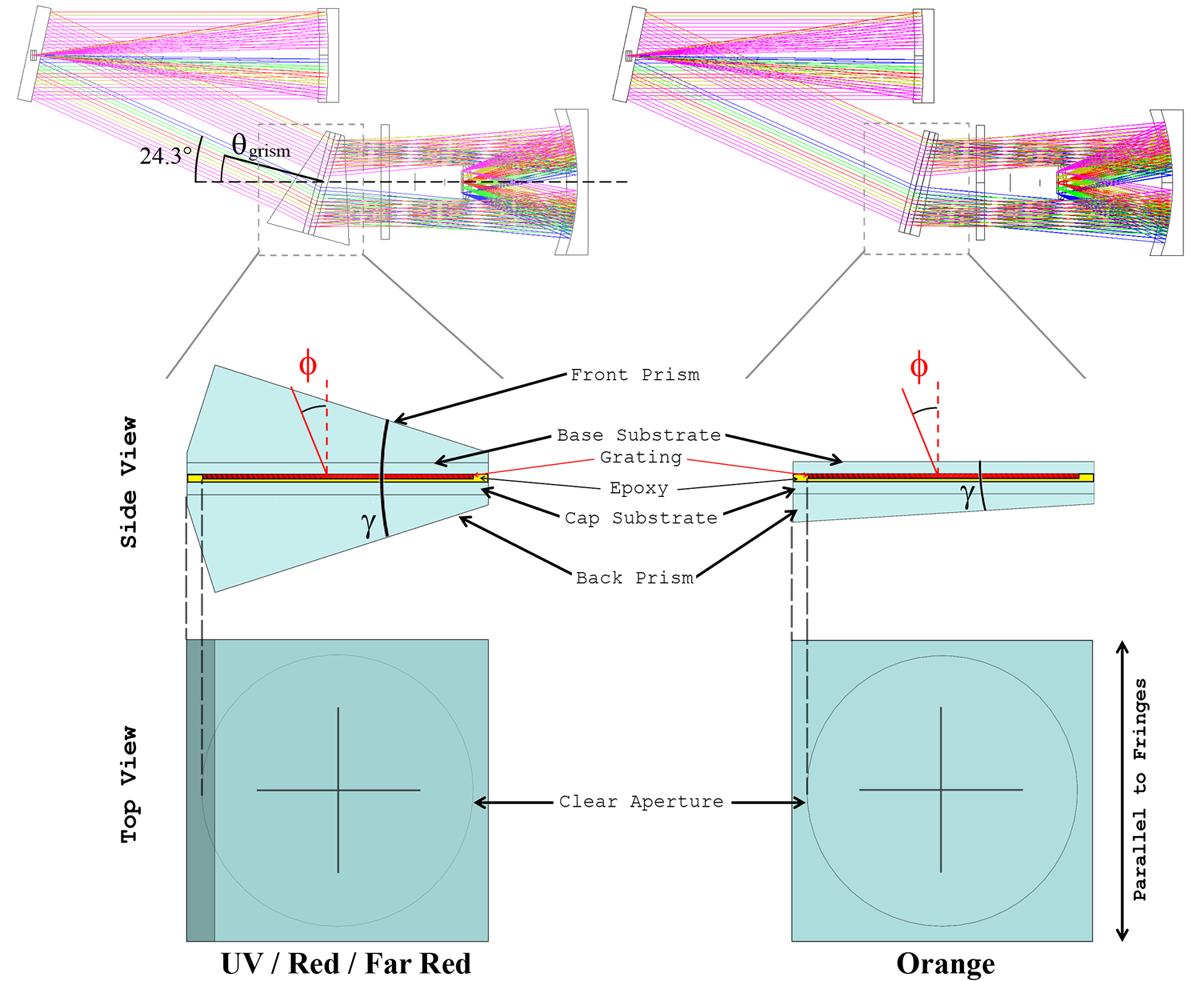}
   \end{tabular}
   \end{center}
   \caption[example] 
   { \label{fig:GrismSchem} 
A schematic of the LRS2 grisms for the UV / Red / Far Red Arms (left column) and Orange Arm (right column). The top row shows a ray-trace of the spectrograph with the grisms in place. Superposed on the UV / Red / Far Red Arm ray-trace is a dashed line that corresponds to the camera optical axis. The solid line is the normal to the VPH diffracting layer. The angle between these, $\theta_{\mathrm{grism}}$, is the grism assembly physical tilt. The middle row shows a side view of each grism with major components labeled. In this view, the fringes are oriented into the page, and the direction of the fringe tilt $\phi$ is indicated. The grating layer thickness is not shown to scale. The grism wedge angle $\gamma$ is also indicated. The bottom row shows a top view of each grism, highlighting the square footprint and circular clear aperture.
}
   \end{figure} 
%------------- 

Using the Bragg condition (e.g., Ref. \citenum{Baldry04}), we can proceed to find the optimal angle of incidence on the grating layer to maximize the diffraction efficiency at the wavelength of our choice (i.e., the Bragg wavelength), which we elect to be near the center of the bandpass for each of the four spectrograph channels. Given the incident angle, the angle of diffraction can be calculated for the central wavelength of the bandpass. The total wedge angle of the grism assembly (labeled $\gamma$ in Fig. \ref{fig:GrismSchem}) can then be determined to match the diffracted beam deviation to the fixed angle of 24.3\degree\ between the collimated beam and the camera optical axis (see Fig. \ref{fig:GrismSchem}). For each channel except the LRS2-B Orange Arm, $\gamma$ is relatively large and requires a prism to be bonded to both sides of the VPH grating (each individual prism has the same wedge angle, i.e., $\gamma/2$). Since the beam deviation for the Orange Arm grating requires that $\gamma$ be a small angle of 3.5\degree, only a single prism is bonded to the backside of the grating to slightly redirect the spectra properly onto the CCD. The physical tilt of the grism assembly (labeled $\theta_{\mathrm{grism}}$ in Fig. \ref{fig:GrismSchem}) is constrained by the need to avoid imaging the Littrow recombination ghost\cite{Burgh07}. As pointed out in Ref. \citenum{Burgh07}, decoupling the Bragg condition from the Littrow configuration is achieved by introducing a tilt to the fringes (labeled $\phi$ in Fig. \ref{fig:GrismSchem}), the amount of which is a free parameter in the grating design and is selected to yield the proper grating layer angle of incidence for meeting the Bragg condition given the constraints of the prism wedge angle and the physical tilt of the grism assembly. The resulting spectral resolution for each channel can be estimated by rearranging and simplifying Eq. A6 from Ref. \citenum{Baldry04} in terms of the VIRUS instrument geometry with a fiber core diameter of 170 $\mu$m. The calculated spectral resolution for each channel is shown in Table \ref{tab:GrismProp} along with additional tabulated parameters that describe the designs of the grisms. The grisms can be seen schematically in Fig. \ref{fig:GrismSchem} along with an example ray-trace of the spectrograph.

\input{t2.tex}

%-------------
   \begin{figure}[t]
   \begin{center}
   \begin{tabular}{c}
   \includegraphics[width=0.9\textwidth]{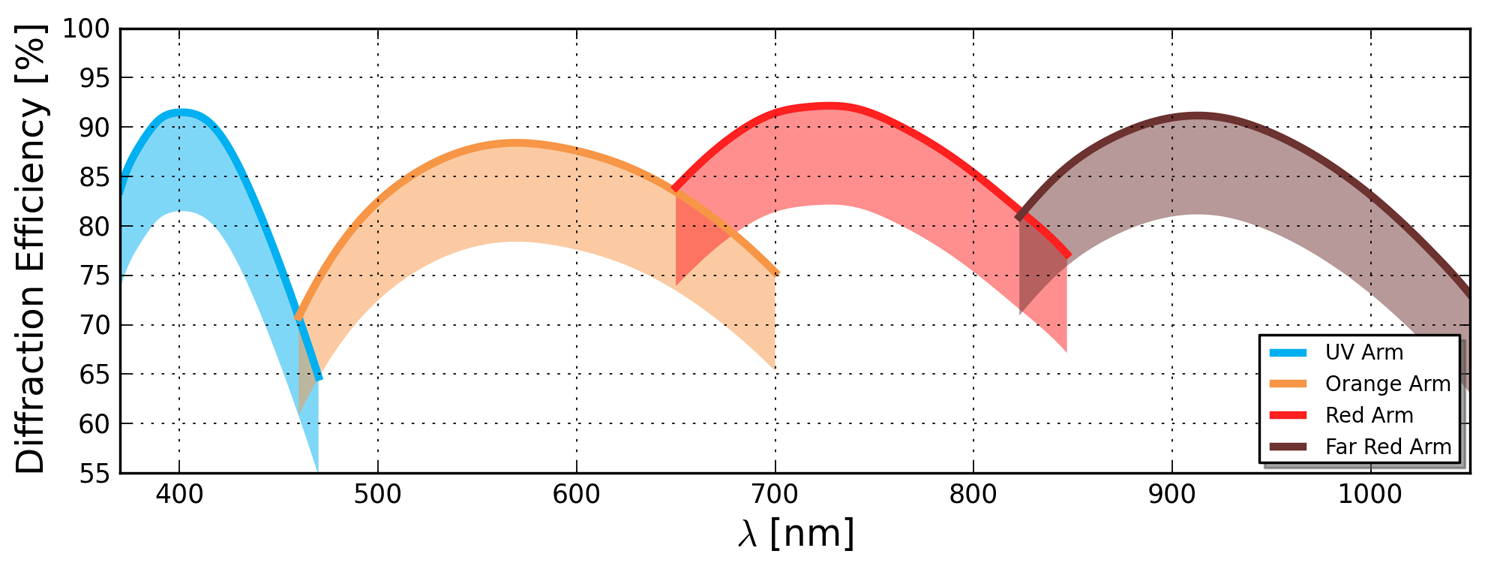}
   \end{tabular}
   \end{center}
   \caption[example] 
   { \label{fig:RCWA} 
RCWA predictions of the first order external diffraction efficiency of the LRS2 grisms. The solid curves are the model predictions taking into account the internal transmittance of the BK7 prisms and substrates, reflection losses, absorption in the epoxy layers, and the transmittance of the DCG layer. The shaded regions beneath each curve represent the range of delivered efficiency that we can tolerate. The parameters describing the modeled DCG layer are tabulated in Table \ref{tab:GrismProp}.
}
   \end{figure} 
%------------- 

The grisms have a square physical footprint with 150 mm sides. The grating layer clear aperture is circular with an over-sized 138 mm diameter as compared to the 125 mm diameter collimated beam to relax the translational position tolerance of the grism within the instrument. We have performed a detailed Monte Carlo analysis of the grism tolerance for each channel including the expected fabrication error, assembly error, and the positional accuracy of the assembly in the instrument. Our criteria for determining an acceptable tolerance level required that no fiber image fall off the CCD, which constrains the image motion to $\pm110$ $\mu$m in the spatial direction. In the spectral direction, we require that the width of overlap region between the two channels in each spectrograph pair be maintained to $\geq50$\% of its original value. In Fig. \ref{fig:dichroic}, we showed the effect of the grism fabrication and alignment tolerance on the position of the channel wavelength limits in the transition region. These spectral variations are dominated by the $\pm2$ line mm$^{-1}$ precision to which the VPH layer fringe density can be fabricated. From this analysis, we have determined that the grisms are straightforward to fabricate and that adequate alignment in the spectrograph only requires a single degree of fine adjustment in rotation about the camera's optical axis. 

In Fig. \ref{fig:RCWA}, we present predictions of the first order external diffraction efficiency of each grism calculated using Rigorous Coupled Wave Analysis (RCWA\cite{Gaylord85}) for unpolarized light. The parameters for the dichromated gelatin (DCG) diffracting layer used in the RCWA models are tabulated in Table \ref{tab:GrismProp}. These theoretical diffraction efficiencies have been corrected to include the internal transmittance of the BK7 prisms and substrates for the average path length through the grism optical assembly, surface reflection losses and absorption in the anti-reflection (AR) coating, absorption in the optical adhesive, and the transmittance of the DCG layer (as scaled from the data in Ref. \citenum{Barden00}). We can accept grisms with diffraction efficiencies no more than 10\% worse than these predictions. Our extensive experience with the VPH-based dispersing elements and feedback from the vendor shows that meeting the lower limit of this requirement is achievable\cite{Hill03,Adams08,Chonis12b,Chonis14}. 

\subsection{Collimator Assembly}\label{subsec:collimator}
\begin{figure}[t]
   \begin{center}
   \begin{tabular}{c}
   \includegraphics[width=0.96\textwidth]{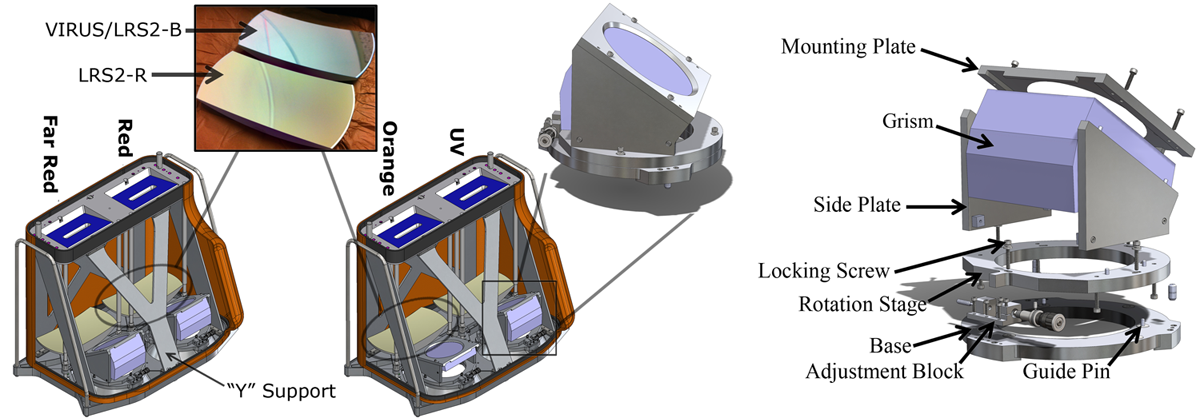}
   \end{tabular}
   \end{center}
   \caption[example] 
   { \label{fig:Collimator} 
Schematic view of the modifications to the VIRUS collimator assembly for LRS2-R and LRS2-B. As indicated in the text, these modification are: 1 - Replacement of the lower support strut to one with a ``Y'' shape to make room for the larger grism dispersers; 2 - For the two LRS2-R channels, replacement of the standard VIRUS reflective optics with custom coated mirrors that have reflectivity optimized for $650-1050$ nm (see the inset photo of the two differently coated mirrors); and 3 - Replacement of the VIRUS VPH diffraction gratings for the grism assemblies, including a custom mount for each channel. At right, an exploded view of the grism mount assembly can be seen with major components labeled. 
}
   \end{figure}

Other than the change from the standard VPH gratings used in VIRUS to larger VPH grisms, the two LRS2 collimator assemblies largely remain optically and mechanically unchanged from the standard VIRUS collimator assembly. The collimator assembly can be seen outlined in Fig. \ref{fig:Collimator}. For all channels except the Orange Arm, the LRS2 grisms are large enough to collide with a support strut on the lower side of the standard VIRUS collimator. This support strut is only for stiffness, and the VIRUS ``V''-shaped part is easily exchanged for a new ``Y''-shaped part to accommodate the larger optics. All of the reflective optics in the LRS2-B collimator assembly utilize the same broadband dielectric coating utilized in VIRUS ($>95$\% average reflectivity for $345-700$ nm, but optimized for $350-590$ nm with $>98$\% average reflectivity). In LRS2-R however, the reflective optics are custom coated with a red optimized dielectric coating ($>98$\% average reflectivity for $650-1050$ nm). With the exception of the grism mounts (discussed below), the construction of the LRS2 collimator assemblies follows directly from the assembly line methods used for VIRUS\cite{Prochaska12,Marshall14,Tuttle14}.

%-------------
%------------- 

The hardware to mount the larger grism dispersers is custom to each of the four LRS2 spectrograph channels. As discussed in the previous subsection, the Monte Carlo optical tolerance analysis for the grism fabrication considered the range of alignment errors that may result from the installation of the assembly into the instrument. This analysis showed that the placement of the grism in the spectrograph can be achieved using mechanical reference features that are positioned with precision machine tolerance. The only fine adjustment necessary for the final grism alignment is to the rotation of the grism assembly about the camera's optical axis. This rotational adjustment serves as a compensator for angular misalignments in the grism assembly (such as the fringe direction relative to external reference features or alignment errors of the prisms relative the fringes) that would otherwise result in the movement of a spectrum of one or more fibers near the edge of the IFU slit off the CCD detector. From 500 Monte Carlo realizations for each grism, we found that the maximum required range of rotational compensation is $\pm0.75$\degree.

The grism mount design can be seen in Fig. \ref{fig:Collimator}. The face of the front prism of the grism assembly is first bonded with RTV outside the clear aperture to a mounting plate that includes a baffle to reduce stray light. The position of the grism on the plate is constrained by machined reference features. Side plates machined specifically for each channel's grism geometry then attach to the mounting plate, and are positioned with precision using locating pins before being fastened. Additional RTV is injected in between the grisms' non-optical faces and the inside of the side plates for additional bonding strength. The side plates attach to a custom rotation stage, and are again positioned with precision using locating pins. Similar to the adjustment mechanism used in the VIRUS grating mounts\cite{Prochaska12}, the rotation is achieved by two guide pins mounted to a base plate sliding within arced slots machined into the rotation stage. Fine adjustment is achieved by a micrometer and opposing spring plunger pushing on a tab that is machined into the rotation stage plate. The rotation stage rides on a series of 6 nylon pads that are attached to the base plate to reduce friction and allow smooth motion. We have designed $\pm2$\degree\ of rotation into the rotation stage assembly so as have ample room for alignment contingency beyond the $\pm0.75$\degree\ prediction from the Monte Carlo analysis. Using the coarse and fine adjustments of the micrometer, a single revolution corresponds to 0.2\degree\ and 0.02\degree\ of rotation, respectively. Once the spectra are aligned on the CCD detector, the rotation stage can be locked into place with a series of 4 locking screws. For additional security after alignment, dabs of epoxy are injected in between the rotation stage and base. The adjustment block (onto which the micrometer and spring plunger are mounted) can be removed from the base allowing only a single adjuster assembly for all four channels to be fabricated. The base plate features the same footprint and locating pin scheme as the VIRUS grating cells to allow the LRS2 grism cell assemblies to fit seamlessly onto the collimator base plates that are fabricated in the VIRUS production.

\subsection{Cameras and Detectors}\label{subsec:camera}
\begin{figure}[t]
   \begin{center}
   \begin{tabular}{c}
   \includegraphics[width=0.98\textwidth]{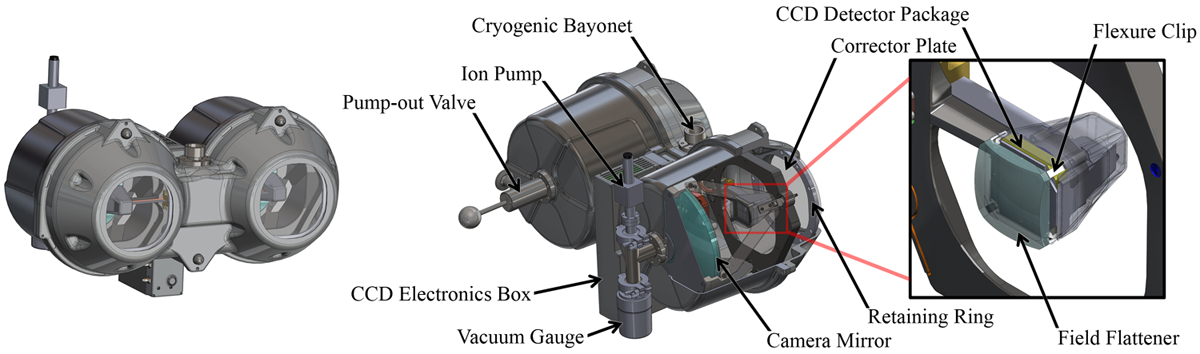}
   \end{tabular}
   \end{center}
   \caption[example] 
   { \label{fig:Camera2} 
Schematic view of the LRS2 cameras. At left is a view of the camera pair looking at the entrance aperture through the Schmidt corrector plates. The CCD detector heads can be seen suspended in front of the camera mirror through each entrance aperture. The middle view shows the camera pair at a different angle with the cast aluminum vacuum chamber cut away to provide a view of the camera internals. Major optical and mechanical components, especially those that are mentioned in the text and are different from the VIRUS camera, are labeled. At right, a zoomed-in view of the CCD detector head is shown.
}
   \end{figure}

The VIRUS camera and detector system design is described in detail in Ref. \citenum{Hill14a}. As with the LRS2 collimator assemblies, only minor modification is needed to convert a VIRUS camera pair for use with LRS2-B or LRS2-R, and the construction of the two LRS2 camera assemblies follows directly from the assembly line methods used for VIRUS\cite{Tuttle14}. An overview of an LRS2 camera pair can be seen in Fig. \ref{fig:Camera2}.

%-------------
%------------- 

\begin{figure}[t]
   \begin{center}
   \begin{tabular}{c}
   \includegraphics[width=0.9\textwidth]{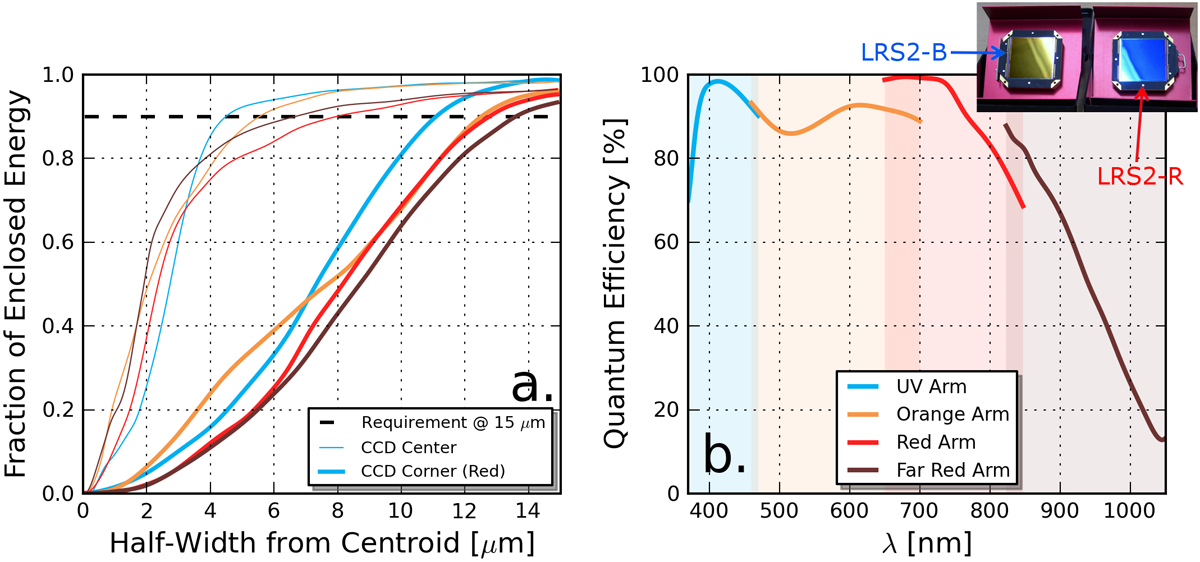}
   \end{tabular}
   \end{center}
   \caption[example] 
   { \label{fig:Camera1} 
\textit{a}) The diffraction ensquared energy within a $2\times2$ pixel area for the four LRS2 spectrograph channels after including the custom transmissive correcting optics in the camera assemblies. As described in the text, the UV Arm utilizes all VIRUS optics, while the Orange Arm utilizes only a custom field flattener. Both LRS2-R channels utilize the same custom corrector plate and field flattener combination that is entirely different from the VIRUS optical prescription. The thin curves correspond to the best image quality at the center of the CCD detector, while the thick curves correspond to the worst image quality which occurs at the corners of the CCD for the reddest wavelengths imaged by each channel. The colors of the curves are as indicated for each spectrograph channel in the panel $b$ legend. \textit{b}) The measured QE of the four LRS2 CCD detectors. Also shown at the top right corner of the plot is a photo of two of the four LRS2 CCDs. The CCD at left (reflecting yellow) is a LRS2-B/VIRUS CCD while the CCD at right (reflecting blue) is a LRS2-R CCD with different AR coating to help boost the QE at longer wavelengths. 
}
   \end{figure}

As mentioned in $\S$\ref{subsec:VIRUS}, the VIRUS camera is optimized for $350 < \lambda\:(\mathrm{nm}) < 550$. While the majority of the optical power is in the fast $f/$1.33 spherical camera mirror, the image quality begins to degrade at wavelengths $>550$ nm due to the change in the index of refraction $\Delta n_{\mathrm{FS}}$ of the fused silica corrector plate and field flattener. For all four LRS2 spectrograph channels, the image quality requirement is such that $\geq90$\% enclosed energy (EE90) is achieved within a $2\times2$ pixel area (i.e., 30 $\mu$m $\times$ 30 $\mu$m) for all wavelengths and all fibers imaged on the CCD detector. Although the fiber spacing on the LRS2 CCDs is less aggressive than for VIRUS, meeting this image quality specification is important to help avoid fiber cross-talk, especially at longer wavelengths where bright night sky emission lines begin to become a factor. For the LRS2-B UV Arm, no departure from the VIRUS camera optical design is required since its wavelength coverage lies entirely within that of VIRUS. While the LRS2-B Orange Arm overlaps with the VIRUS wavelength range, it does extend beyond it by 150 nm. Using VIRUS optics in the Orange Arm, wavelengths $>680$ nm in the 20\% of fibers that are farthest from the center of the fiber slit drop below the EE90 requirement. We corrected this by simply changing the even aspheric surface on the rear side of the field flattener optic and reducing its overall thickness while maintaining the use of the VIRUS corrector plate. Since the distance from the detector surface to the vertex of the new aspheric surface of the field flattener is the same as for VIRUS, no change in the mounting hardware is required. For the LRS2-R channels, the image quality is poor enough with the VIRUS corrective camera optics that entirely new prescriptions for both the field flattener and corrector plate are required. However, $\Delta n_{\mathrm{FS}}$ is $\sim2\times$ smaller over the wavelength range covered by both LRS2-R channels combined than for either of the LRS2-B channels individually. This suggests that a single prescription may be able to be found for the camera correcting optics for the Red and Far Red channels together. Indeed, we have derived a single combination of field flattener and corrector plate that yield excellent image quality for the entire LRS2-R wavelength range. For LRS2-R, the fused silica corrector plate has a different aspheric surface and is thinner at the edge by 0.3 mm as compared to VIRUS. Since the corrector plate also serves as the vacuum window for the camera cryostat, we have modified the aluminum retaining ring that provides the O-ring vacuum seal to accommodate the thinner optic and ensure a good vacuum seal. The LRS2-R field flattener also has a different aspheric prescription and is thinner at the edge as compared to VIRUS. The smaller edge thickness of the field flattener requires four custom Invar flexure clips for each channel to be fabricated for mounting the optic on top of the CCD detector. As with the LRS2-R collimator reflective optics, the camera mirrors for the two LRS2-R channels are coated with the same high reflectivity dielectric coating that is optimized for $650-1050$ nm. Similarly, the LRS2-R transmissive camera correcting optics also have custom AR coatings that are optimized for this wavelength range. The image quality of all four LRS2 channels after including the custom camera correcting optics is shown in Fig. \ref{fig:Camera1}$a$.

%-------------
%------------- 

The LRS2-B cameras utilize the same custom $2064\times2064$ CCD detectors as VIRUS. The detectors are thinned, backside illuminated CCDs with 15 $\mu$m square pixels and AR coatings that are optimized for the UV-blue. The CCDs are packaged specifically for the VIRUS camera design\cite{Hill14a}. The CCDs feature high quantum efficiency (QE) in the UV-blue, and low read-noise of $<3.1$ $e^{-}$. The detector system is being supplied by Astronomical Research Cameras, Inc., and can read the pair of CCDs binned $2\times1$ in $\sim20$ seconds. Fig. \ref{fig:Camera1}$b$ shows the measured QE of the LRS2-B CCDs. For LRS2-R, custom CCDs with identical physical and electrical format to the VIRUS CCDs are utilized, which greatly reduces the overall cost of the LRS2-R spectrograph by avoiding a significant redesign of the camera's detector mounting head to utilize a different type of CCD package. The CCDs have custom AR coatings and a thicker epitaxial layer of 30 $\mu$m to help mitigate charge diffusion effects, boost sensitivity at longer wavelengths, and reduce fringing. The LRS2-R CCDs feature low read-noise of $<3.7$ $e^{-}$, and their QE can also be seen plotted in Fig. \ref{fig:Camera1}$b$. The LRS2 CCDs are shuttered externally by the same large shutter for VIRUS that is built in front of the PFIP focal surface (see $\S$\ref{sec:operation}).

By using the same cameras and detector system as VIRUS, we are able to utilize the design and investment made in its cryostat design and its infrastructure for cryogenics\cite{Chonis10} and CCD readout\cite{Tuttle14}. For VIRUS, significant cost savings were realized by not outfitting the camera pairs with vacuum measurement equipment. VIRUS will undergo a maintenance schedule in which each camera pair is warmed, pumped, and then re-cooled in-situ every 3-4 months, which is approximately the amount of time verified in the lab that a camera can maintain an adequately low pressure for the CCDs to remain at operational temperature\cite{Hill14a}. Unlike VIRUS, the LRS2-B and LRS2-R camera pairs will be outfitted with full range vacuum gauges and ion pumps to ensure the longevity of the vacuum since we wish to avoid down time. Experience with LRS indicates that the vacuum can be maintained for $>18$ months as long as the instrument remains cold.
%%%%%%%%%%%%%%%%%%%%%%%%%%%%%%%%%%%%%%%%%%%%%%%%%%%%%%%%%%%%%

%%%%%%%%%%%%%%%%%%%%%%%%%%%%%%%%%%%%%%%%%%%%%%%%%%%%%%%%%%%%%
\section{DEPLOYMENT \& OPERATION}\label{sec:operation}
\subsection{Final Alignment of the Spectrographs}\label{subsec:alignment}
The LRS2 collimators and cameras are assembled using the standard assembly line practices established for VIRUS in Refs. \citenum{Tuttle14,Marshall14} and \citenum{Prochaska12}. Once the collimator and camera for an LRS2 spectrograph pair are assembled, roughly aligned, and integrated together with the IFU, the final adjustment of the camera mirror is made to achieve the best focus over the full field. This is accomplished using full-field image moment-based wavefront sensing\cite{Lee12b,Lee12c,Lee14a}, where the camera mirror is actuated through focus and CCD images are taken in discrete steps with an arc lamp illuminating the IFU. The final adjustment to the camera mirror position is rapidly derived from the analysis of the resulting set of images. The adjustment is performed with the detectors at operational cryogenic temperatures and utilizes a specially designed temporary vacuum housing with feed-throughs to actuate and clamp the camera mirror into its final position. The final optical alignment of the LRS2 spectrographs should not require further adjustment at the telescope once deployed since the instruments operate in a constant gravity vector (see below) and were specifically designed to be extremely stable against temperature changes through the use of Invar in the critical places within the mechanical design\cite{Hill14a}.

\subsection{LRS2-B and LRS2-R at the HET}\label{subsec:deployment}
By utilizing the VIRUS design as the basis of the two LRS2 spectrograph pairs, we are able to take advantage of the large amount of infrastructure that is installed at the HET to support the VIRUS spectrograph array. LRS2-B and LRS2-R will be rack mounted inside of the large environmentally controlled enclosures\cite{Prochaska14} of the VIRUS Support Structure\cite{Good14b} that is mounted to the sides of the HET telescope truss structure (see Fig. \ref{fig:FiberCable1}). They will be located at the top of the enclosure as close to the PFIP as possible to keep the overall fiber cable length short. The enclosure ports for LRS2 will also contain some additional infrastructure that is not required for a VIRUS spectrograph pair (e.g., AC power and electronics racks for mounting the hardware and controllers for the vacuum equipment). The VIRUS Support Structure also supports the extensive liquid nitrogen distribution system\cite{Smith08,Chonis10} that is utilized to keep the CCDs contained in the 75 VIRUS spectrograph pairs and the two LRS2 spectrograph pairs cool. At the top of each enclosure pair is a header tank, which provides a continuous flow of cryogen to the various lateral levels of vacuum jacketed pipes. The header tanks are replenished from a large external dewar, which is refilled by an 18-wheel tanker truck every two weeks. From each of the lateral sections of pipe in each enclosure are three vacuum jacketed flexible hoses that feed cryogen to each spectrograph pair. The flexible hoses terminate with a novel copper heat exchanger bayonet\cite{Chonis10} that conducts heat out of the camera cryostat and allows the camera to be disconnected from the cryogenic system with ease without affecting the rest of the system.

The two LRS2 IFU feeds are integrated near the center of the HET field of view on the same focal plane assembly on which all of the VIRUS IFUs are mounted. This allows the implementation of parallel observing\cite{Odewahn12} where VIRUS can observe in the background while one of the LRS2 spectrograph pairs are observing their target. This configuration requires that LRS2 and VIRUS share the same shutter system, which is located within the PFIP in front of the fibers\cite{Vattiat14}. LRS2-B and LRS2-R are operated independently and the detectors are read out by a dedicated computer that uses the same data system architecture as VIRUS. However, the VIRUS data acquisition system and the LRS2 control computer must communicate when parallel mode is being utilized to coordinate shutter open and close times and the three dither moves that are required to fill in the VIRUS IFU field of view. The VIRUS dithers are extremely precise\cite{Vattiat14}, so the movement of the field on the LRS2-B or LRS2-R IFU between the three separate exposures can be communicated back to the data reduction software to properly reconstruct the final spectra. When not using parallel mode, LRS2-B and LRS2-R can be utilized independently depending on the science goals since the two instruments observe two different fields that are separated by $\sim100$\arcsec. Additionally, if an object is intended to be observed with the full LRS2 wavelength range (i.e., $370-1050$ nm), one can simply beam-switch between the LRS2-B and LRS2-R IFU fields to observe the target object while the other spectrograph pair is obtaining sky spectra. Calibration of the LRS2 data are provided within the PFIP by the Facility Calibration Unit\cite{Lee12d}, which is a deployable optical assembly that mimics the illumination pattern of the HET primary mirror and injects suitable calibration light for obtaining flat fields and wavelength calibration for any of the HET facility instruments into the WFC.

All of the above operations will be conducted and controlled through the HET queue scheduling system\cite{Shetrone07} to ensure that optimal observations are obtained in response to weather conditions, high priority targets, TOOs, and to advance the completion of observing programs. The main operational change between LRS2 and the current LRS is from slit spectroscopy to IFU observations. The IFU, in addition to the improved metrology\cite{Lee12a} provided through the HET WFU, should reduce target acquisition to $\sim1$ minute from the current value of $\sim10-15$ minutes. Together with the fact that the two LRS2 spectrograph pairs have no moving parts, we expect the operation of LRS2 to be extremely reliable and efficient.

\subsection{Data Reduction}\label{subsec:reduction}
A data reduction pipeline will be provided for reducing the LRS2 integral field data based on the CURE package that has been developed for VIRUS\cite{Snigula12}. Similar to the adaptation of the VIRUS hardware that we have outlined here for transforming VIRUS pairs into LRS2, CURE will only require minor adaptations to account for the different spectral data format of LRS2 (i.e., wavelength range, the number of fibers, the fiber spacing, etc.). The pipeline will provide basic reduction and calibration of the data cubes and provide sky subtraction routines that are tailored to the types of expected targets (i.e., unresolved, resolved but smaller than the IFU, and larger than the IFU for which beam-switching will be required to sample the sky). The data cubes can be visualized and analyzed with CURE or with other IFU-specific packages, such as P3D\cite{Sandin10}.
%%%%%%%%%%%%%%%%%%%%%%%%%%%%%%%%%%%%%%%%%%%%%%%%%%%%%%%%%%%%%

%%%%%%%%%%%%%%%%%%%%%%%%%%%%%%%%%%%%%%%%%%%%%%%%%%%%%%%%%%%%%
\section{SENSITIVITY MODEL}\label{sec:sensitivity}
\subsection{Technical Description}\label{subsec:sensdescription}
\begin{figure}[t]
   \begin{center}
   \begin{tabular}{c}
   \includegraphics[width=0.98\textwidth]{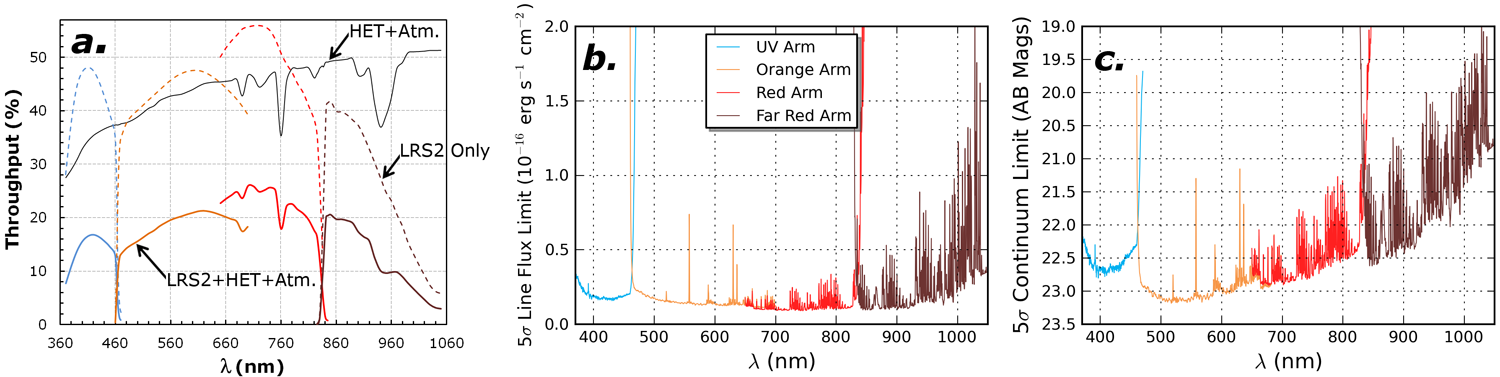}
   \end{tabular}
   \end{center}
   \caption[example] 
   { \label{fig:Sensitivity} 
The calculated sensitivity of the LRS2 spectrographs. \textit{a}) The throughput of the instrument using the most recent data on all of the optical components. The solid black curve shows the combined throughput of the HET and the sky at an airmass of 1.22, but does not include atmospheric or Galactic extinction or account for the mean illumination of the HET primary mirror during a track. The dashed colored curves are the throughput of the four LRS2 spectrograph channels, while the solid colored curves give the total sky-to-detector throughput. Utilizing this sky-to-detector throughput, panel $b$ shows the one-dimensional $5\sigma$ line flux limit for the LRS2 channels in a 20 minute exposure with 1.2\arcsec\ FWHM seeing in dark sky conditions, assuming a 100 km s$^{-1}$ FWHM intrinsic line width. We also assume that the target is unresolved and is centered on one of the IFU spatial elements. \textit{c}) The $5\sigma$ continuum limit in AB magnitudes for the same conditions.\vspace{5mm}
}
   \end{figure}

Refs. \citenum{Lee10} and \citenum{Chonis12a} presented initial predictions of the throughput of LRS2. These predictions were were based on the throughput model used for VIRUS, which has been verified against on-sky tests using the Mitchell Spectrograph\cite{Hill08b}. For LRS2, additional factors must be taken into account for the transmission efficiency of the IFU feed optics and the lenslet coupling to the fibers. Here, we update the initial LRS2 throughput predictions to include the as-built efficiency measurements for all of the components for which we have taken delivery. At the time of writing, we have received all optical components for the four LRS2 spectrograph channels except the VPH grisms. As such, we assume the lower limit of the acceptable range of the modeled external diffraction efficiency for the grisms as shown in Fig. \ref{fig:RCWA}. In Fig. \ref{fig:Sensitivity}$a$, we show the newly calculated expected throughput of LRS2 both with and without the combined effect of the HET and the atmosphere, which is shown by the black curve. Our calculation of the ``HET + atmospheric'' throughput takes into account the most recent measurements of the HET primary mirror's mean reflectivity, the throughput of the WFC, and the transmission of the atmosphere at an airmass of 1.22 (which is constant for the purposes of these calculations given the HET's fixed altitude design). While the throughput shown in Fig. \ref{fig:Sensitivity}$a$ does not take into account atmospheric extinction, Galactic extinction, or the mean illumination of the HET's primary mirror during a track, these factors can easily be added if desired.

%-------------
%------------- 
\input{t3.tex}

To aid in the development of the initial science programs that LRS2 will undertake, the above expected throughput is used as input for a tool that we have developed to simulate an LRS2 observation and output the expected wavelength dependent instrumental sensitivity. For a spatially unresolved target, the script simulates a two-dimensional Moffat point spread function (PSF) at the HET focal plane with a given FWHM to simulate various seeing conditions. An array of apertures simulating the LRS2 IFU is overlaid on the PSF image, and the relative fluxes in each IFU spatial element are calculated. The IFU aperture array can be moved relative to the PSF to simulate different centration scenarios, which can affect the final SNR of the observation. Given a desired exposure time, the noise in each IFU spatial element is calculated using the measured read noise of the CCDs, the size of a single spectral resolution element in pixels, and the sky flux. The sky flux in each IFU spatial element is calculated by convolving the measured sky spectra from Ref. \citenum{Hanuschik03} to each LRS2 channel's spectral resolution and dispersion and scaling it to the typical dark sky conditions at McDonald Observatory (here, we assume a $B$-band sky surface brightness of 22.5 mag arcsec$^{-2}$). Given a desired significance level per spectral resolution element, the script determines which of the IFU spatial elements would have enough object flux to make a positive contribution to the overall SNR, and the total noise per spectral resolution element from those spatial elements is calculated. To give the flux limit at the chosen significance level for a one-dimensional combined spectrum, the total noise per resolution element calculated at that wavelength is then multiplied by the significance level, converted to sensible flux units, and corrected for any flux that was missed in individual IFU spatial elements due to insufficient SNR. For a resolved object, the total noise per spectral resolution element in a single IFU spatial element is multiplied by the significance level, converted to sensible flux units, and finally divided by the on-sky area covered by the lenslet to provide a surface brightness (SB) limit. The wavelength dependent sensitivity can be output both in terms of a continuum flux limit or a line flux limit. The line flux limit can be calculated for an emission line of a chosen intrinsic width.

As an example of LRS2's capabilities, we have calculated the $5\sigma$ line flux limit and the 5$\sigma$ continuum limit for a 20 minute exposure with the CCD binned $2\times1$ in the spectral direction with the four LRS2 spectrograph channels on the HET in panels $b$ and $c$ of Fig. \ref{fig:Sensitivity}, respectively. For these calculations, we assume that an unresolved object is observed in 1.2\arcsec\ FWHM seeing conditions with the target centered perfectly on one of the IFU spatial elements. Since the simulated seeing conditions result in the object's light being spread over multiple IFU spatial elements, the flux limits reflect the sensitivity of a coadded one-dimensional spectrum. For the line flux limit calculation, an intrinsic emission line width of 100 km s$^{-1}$ FWHM is assumed. In addition to these sensitivity predictions, Table \ref{tab:ExpTime} shows the minimum exposure time required for the sky noise to be equal to the detector read noise in a single spectral resolution element for select wavelengths in each of the four LRS2 spectrograph channels. The sky flux used in the calculation is taken as the median over the $\pm50$ \AA\ range centered on each listed wavelength.

\subsection{Application: \lya\ Blobs at $2<z<3$}\label{subsec:sensapplication}
As a specific example of LRS2's capabilities, we apply the sensitivity predictions above to the science case of studying \lya\ nebulae (or, \lya\ Blobs; LABs). LABs are giant structures of HI \lya\ emitting gas observed in the high redshift universe ($z > 2$), typically having sizes of up to 100 kpc and \lya\ luminosities of $10^{43} - 10^{44}$ erg s$^{-1}$. These objects are rare, but large area surveys are now discovering dozens at a time (e.g., Ref. \citenum{Yang10}). LABs appear to be highly clustered and tend to be discovered in overdense regions, suggesting that they are associated with the formation of massive galaxies or clusters in some of the most massive dark matter halos. Recent work suggests that LABs trace large-scale filamentary structure\cite{Erb11} and that their morphology depends strongly on the environment in which they live\cite{Matsuda12}. The origin of LABs is not well understood, but three main mechanisms for powering the extended \lya\ emission have been proposed: 
\begin{enumerate} \itemsep1pt \parskip0pt \parsep0pt
   \item Cooling radiation from $\sim10^{4}$ K gas that is being accreted into massive dark matter halos (e.g., Ref. \citenum{Haiman00}) 
   \item Shock induced heating of gas by mechanical feedback from galactic superwinds (e.g., Ref. \citenum{Taniguchi00}) 
   \item Photoionization of gas by a central starburst or AGN (e.g., Refs. \citenum{Hayes11,Geach09}) 
\end{enumerate}
The latter two mechanisms are attractive explanations given that most LABs contain at least one embedded optical, IR, or X-ray detected source. However, some LABs contain no detectable central power source, leaving the first mechanism as the most likely scenario\cite{Nilsson06}. While all three mechanisms could be at play simultaneously for a given LAB, it is clear that these objects are sites of strong interaction between galaxies and the intergalactic medium and depend on the distribution and kinematics of the gas that lives in these massive dark matter halos. Due to the resonant scattering nature of the \lya\ transition, the shape of the emergent \lya\ spectral line profile is sensitive to the column density and velocity field of neutral gas along the line-of-sight (e.g., Refs. \citenum{Verhamme06,Barnes11}). While interpretation of the \lya\ line profile is non-trivial, detailed radiative transfer modeling of spatially resolved \lya\ spectra has been shown to provide strong constraints on the gas distribution and velocity field\cite{Adams09}. Combined with constraints on the physical nature of the galaxies that may be embedded in LABs, spatially resolved observations aimed at studying how the \lya\ emission is distributed and how its spectral line profile varies as a function of position across the LAB can further our understanding of the relative contributions of each powering mechanism. To date, only four LABs have been studied with such integral field observations\cite{Adams09,Bower04,Wilman05,Weijmans10,Francis13}. With LRS2 (the UV Arm, specifically), similar observations of LABs existing in different environments (e.g., field vs. overdensity) and containing different embedded sources (e.g., AGN vs. starburst vs. none) could be carried out systematically for the first time.

%-------------
   \begin{figure}[t]
   \begin{center}
   \begin{tabular}{c}
   \includegraphics[width=0.98\textwidth]{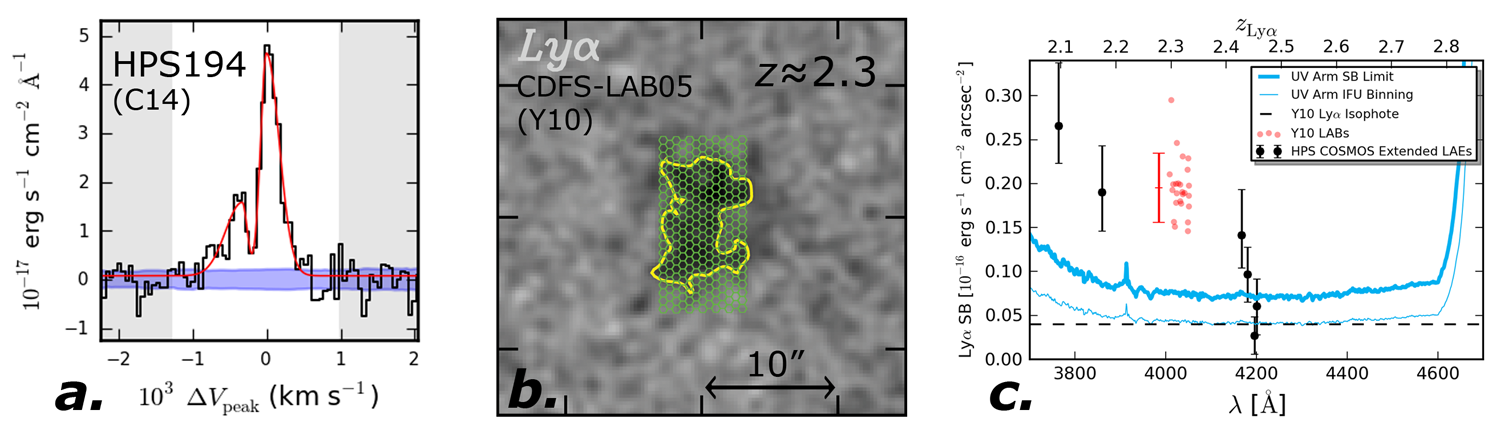}
   \end{tabular}
   \end{center}
   \caption[example] 
   { \label{fig:LyA} 
Observing LAEs and LABs with LRS2. \textit{a}) A 1D slit spectrum of HPS194, a $z\approx2.3$ LAE discovered in the HPS\cite{Adams11}. The spectrum consists of 5.25 hours of exposure from Ref. \citenum{Chonis13b} and was taken with the IMACS\cite{Dressler11} spectrograph on the 6.5 m Magellan Baade telescope at 2.2 \AA\ FWHM spectral resolution (i.e., the same as the LRS2-B UV Arm). The blue shaded region represents the $\pm1\sigma$ uncertainty, and the red curve is a functional fit to the \lya\ line profile and is used to quantify the line shape. The width of a \lya\ emission component is typically $\sim300$ km s$^{-1}$. \textit{b}) A narrowband \lya\ image of CDFS-LAB05 from Ref. \citenum{Yang10}, a $z\approx2.3$ LAB. The yellow isophote corresponds to a \lya\ SB of $4\times10^{-18}$ \sbcgs. We have also overlaid the LRS2 IFU on the image showing that the field size is well matched to the typical sizes of LABs at $2\lesssim z \lesssim 3$. \textit{c}) A plot of the expected LRS2-B UV Arm $5\sigma$ \lya\ SB limit for 3 hours of exposure. The details of the calculation are described in the text. We also show the deeper limit one obtains when binning 3 IFU spatial elements together (thin curve). For reference, the dashed black horizontal line indicates the depth of the yellow isophote in panel $b$. The black data points indicate the estimated \lya\ SB of LAB candidates in the COSMOS field from the HPS\cite{Adams11}. Red data points correspond to the LAB sample from Ref. \citenum{Yang10}, referred to as ``Y10'' in the legend. The red error bar shows the typical uncertainty on the average \lya\ SB of the ``Y10'' sample.  
}
   \end{figure} 
%------------- 

The LRS2-B UV Arm will be well suited to follow-up studies of LABs and candidate extended \lya\ emitters that have been discovered in the HPS\cite{Adams11} and will be discovered in HETDEX\cite{Hill08a}. The UV Arm covers \lya\ for $2.0 \lesssim z \lesssim 2.9$ with sufficient spectral resolution for resolving the \lya\ line (e.g., Ref. \citenum{Chonis13a,Chonis13b}; see Fig. \ref{fig:LyA}$a$), which is required for its interpretation and comparison to radiative transfer simulations. Objects within this redshift range also have their rest-frame optical emission lines (e.g., \hb, \oiii, \ha, \nii) redshifted into the atmospheric windows in the near-IR, which is essential for characterizing embedded galaxies. Additionally, UV emission lines that are useful for determining the ionization structure of the LAB (e.g., CIV$\lambda1549$ and HeII$\lambda1640$; Ref. \citenum{Scarlata09}) would fall within the LRS2-B Orange Arm's spectral window and are observed simultaneously with \lya\ by the UV Arm. Finally, the LRS2 IFU is well matched in size to the typical LAB at $z\approx2.4$ (104 kpc $\times$51 kpc; see Fig. \ref{fig:LyA}$b$).

In Fig. \ref{fig:LyA}$c$, we show the UV Arm $5\sigma$ \lya\ SB limit per resolution element under dark sky conditions for a 300 km s$^{-1}$ FWHM emission line after $6\times1800$ second exposures. We include 0.2 mag of atmospheric extinction, galactic extinction consistent with a typical extragalactic survey field ($A_{V}\approx0.05$ mag), and assume $2\times1$ CCD binning along the spectral direction. Additionally, we assume that the object is located in an equatorial field such that the 10 m diameter HET pupil sweeps across its primary mirror with an average illumination factor of 75\%. Under these conditions, the UV Arm reaches a median $5\sigma$ SB limit of $7.8\times10^{-18}$ \sbcgs\ while the Orange Arm, which is not plotted in Fig. \ref{fig:LyA}$c$ but collects data congruently with the UV Arm, reaches $4.8\times10^{-18}$ \sbcgs. To show that LRS2-B's UV Arm is capable of mapping the \lya\ emission of LABs under these realistic conditions, we have also plotted the spatially averaged \lya\ SB measurements for the sample of $z\approx 2.3$ LABs from Ref. \citenum{Yang10}, which range in size from $10-60$ arcsec$^{2}$. One such LAB is shown in Fig. \ref{fig:LyA}$b$, and its extent is highlighted by the faint $4\times10^{-18}$ \sbcgs\ isophote that would fall mostly within the LRS2 IFU field with room to spare for simultaneous sky sampling. By taking a slight reduction in spatial resolution in the dimmer regions and binning 3 contiguous IFU spatial elements together, our calculations in Fig. \ref{fig:LyA}$c$ show that this faint isophote can be reached at the $5\sigma$ level with the UV Arm in 3 hours. In addition to the sample from Ref. \citenum{Yang10}, we also show in Fig. \ref{fig:LyA}$c$ the estimated \lya\ SB of six LAB candidates discovered in the 71.6 arcmin$^{2}$ area of the COSMOS field that was surveyed as a part of the HPS\cite{Adams11}. The characteristics of this small LAB sample are diverse, including four LABs that are located in a known overdensity at $z\approx2.45$ and two located in the field. Of the six, two LABs also have associated X-ray detections (suggesting AGN photoionization as the dominant powering mechanism), while the remaining four do not. During HETDEX operations, the VIRUS\cite{Hill14a} instrument will survey an area larger than 71.6 arcmin$^{2}$ in $<1$ hour, likely yielding many dozens of new LAB candidates each night. This provides an excellent opportunity to amass an enormous catalog of extended \lya\ emitting objects with a diverse set of properties for which integral field \lya\ spectra can be obtained with LRS2, which will yield the first systematically observed LAB sample not preselected to lie in known overdensities. 
%%%%%%%%%%%%%%%%%%%%%%%%%%%%%%%%%%%%%%%%%%%%%%%%%%%%%%%%%%%%%

%%%%%%%%%%%%%%%%%%%%%%%%%%%%%%%%%%%%%%%%%%%%%%%%%%%%%%%%%%%%%
\section{STATUS \& OUTLOOK}\label{sec:status}
As has been described in this paper, LRS2 will be a worthy replacement for the workhorse LRS instrument for the HET. With the new spectroscopic and imaging capabilities coming on line through the HET WFU\cite{Hill14b}, LRS2 will add broadband integral field spectroscopic capability in a robust package. As a capable general purpose instrument for survey follow-up studies, LRS2 will help further the HET's goal of becoming the premier spectroscopic survey telescope. All aspects of the LRS2 design are complete, and we have taken delivery of most of the hardware required for both LRS2-B and LRS2-R. Both spectrograph pairs are currently being built using the assembly standards adopted throughout the mass production of VIRUS, and we expect to complete lab integration tests with the completed IFU assemblies by the fall of 2014. First light of the LRS2 spectrographs is currently paced by the HET WFU schedule\cite{Hill14b}, so both LRS2-B and LRS2-R will be available for commissioning together as soon as they can be accepted at the telescope. 

Along with the LRS2-specific sensitivity model that we have developed, an initial science team has been assembled to help develop commissioning projects and larger initial science programs to push the first LRS2 datasets toward quick publication. These include, but are not limited to:
\begin{itemize} \itemsep1pt \parskip0pt \parsep0pt
   \item A systematic integral field survey of LABs (see $\S$\ref{subsec:sensapplication})
   \item Integral field studies of evolving spiral galaxies at $z<0.3$ 
   \item Follow-up of high proper motion objects to identify substellar members of the galactic halo 
   \item Emission line diagnostics of extremely metal-poor galaxies found with HETDEX at $z<0.1$
   \item Spectroscopic redshift confirmation of galaxies out to $z\gtrsim7$
   \item Synoptic studies of quasar broad absorption line variability events
\end{itemize}
These initial programs will provide verification of the hardware and the software pipeline, encourage continuing refinement of the instrument and methods to improve LRS2's performance for more challenging future science goals, and will highlight the new capabilities of both the upgraded HET and its new low resolution facility spectrograph.
%%%%%%%%%%%%%%%%%%%%%%%%%%%%%%%%%%%%%%%%%%%%%%%%%%%%%%%%%%%%%

%%%%%%%%%%%%%%%%%%%%%%%%%%%%%%%%%%%%%%%%%%%%%%%%%%%%%%%%%%%%%
\acknowledgments     %>>>> equivalent to \section*{ACKNOWLEDGMENTS}       
HETDEX is run by the University of Texas at Austin McDonald Observatory and Department of Astronomy with participation from the Ludwig-Maximilians-Universit\"{a}t M\"{u}nchen, Max-Planck-Institut f\"{u}r Extraterrestriche-Physik (MPE), Leibniz-Institut f\"{u}r Astrophysik Potsdam (AIP), Texas A\&M University (TAMU), Pennsylvania State University, Institut f\"{u}r Astrophysik G\"{o}ttingen (IAG), University of Oxford, and Max-Planck-Institut f\"{u}r Astrophysik (MPA).  In addition to Institutional support, HETDEX is funded by the National Science Foundation (grant AST-0926815), the State of Texas, the US Air Force (AFRL FA9451-04-2-0355), the Texas Norman Hackerman Advanced Research Program under grants 003658-0005-2006 and 003658-0295-2007, and generous support from private individuals and foundations.

We thank the staffs of McDonald Observatory, AIP, MPE, TAMU, Oxford University Department of Physics, and IAG for their contributions to the development of VIRUS. T.S.C. acknowledges the support of a National Science Foundation Graduate Research Fellowship. 
%%%%%%%%%%%%%%%%%%%%%%%%%%%%%%%%%%%%%%%%%%%%%%%%%%%%%%%%%%%%%

%%%%% References %%%%%
\bibliography{ms_arXiv}         %>>>> bibliography data in report.bib
\bibliographystyle{spiebib}   %>>>> makes bibtex use spiebib.bst
%%%%%%%%%%%%%%%%%%%%%%

\end{document}

%% file: t1.tex
\begin{table}[b!]
\caption{LRS2 Feed Optics and IFU Properties} 
\label{tab:IFUProp}
\begin{center}       
\begin{tabular}{|c|c|c|c|c|c|c|} 
\hline
\rule[-1ex]{0pt}{3.0ex} \footnotesize{\textbf{Dichroic}} & \footnotesize{\textbf{Feed}} & \footnotesize{\textbf{Lenslet}} & \footnotesize{\textbf{Micro-Image}} & \footnotesize{\textbf{Fiber Core}} & \footnotesize{\textbf{Lenslet}} & \footnotesize{\textbf{Spaxel Center-to-}}  \\

\rule[-1ex]{0pt}{3.0ex} \footnotesize{\textbf{Crossover}} & \footnotesize{\textbf{Focal}} & \footnotesize{\textbf{Pitch}} & \footnotesize{\textbf{Diameter}} & \footnotesize{\textbf{Diameter}} & \footnotesize{\textbf{Array}} & \footnotesize{\textbf{Center IFU Size}} \\

\rule[-1ex]{0pt}{3.5ex} \footnotesize{(nm)} & \footnotesize{\textbf{Ratio}} & \footnotesize{(\arcsec)} & \footnotesize{($\mu$m)} & \footnotesize{($\mu$m)} &  \footnotesize{\textbf{Format}} & (\arcsec) \\
\hline
\rule[-1ex]{0pt}{0.8ex}  & & & & & & \\
\rule[-1ex]{0pt}{1.8ex}  \footnotesize{463.9 (LRS2-B)} & \footnotesize{$f$/10.5} & \footnotesize{0.59 (0.30 mm)} & \footnotesize{120} & \footnotesize{170} & \footnotesize{$22\times13$} & \footnotesize{$12.4\times6.1$} \\
\rule[-1ex]{0pt}{1.8ex}  \footnotesize{835.0 (LRS2-R)} & & \footnotesize{Hexagonal} & & & \footnotesize{Rectangular} & ($6.3\times3.1$ mm) \\
\hline
\end{tabular}
\end{center}
\end{table}

%% file: t2.tex
\begin{table}[b!]
\caption{LRS2 VPH Grism Properties} 
\label{tab:GrismProp}
\begin{center}       
\begin{tabular}{|r|c|c|c|c|l|} 
\hline
\rule[-1ex]{0pt}{3.0ex} \small{} & \small{\textbf{UV Arm}} & \small{\textbf{Orange Arm}} & \small{\textbf{Red Arm}} & \small{\textbf{Far Red Arm}} & \small{Units (Comments)} \\
\hline
\rule[-1ex]{0pt}{2.0ex}  \scriptsize{$\lambda_{\mathrm{min}}$, $\lambda_{\mathrm{max}}$} & \scriptsize{370, 470} & \scriptsize{460, 700} & \scriptsize{650, 847} & \scriptsize{823, 1050} & \scriptsize{nm} \\
\hline
\rule[-1ex]{0pt}{2.0ex}  \scriptsize{Dispersion} & \scriptsize{0.48} & \scriptsize{1.16} & \scriptsize{0.95} & \scriptsize{1.10} & \scriptsize{\AA\ px$^{-1}$ ($1\times1$ binning)} \\
\hline
\rule[-1ex]{0pt}{2.0ex}  \scriptsize{Spectral Resolution $\delta\lambda$} & \scriptsize{2.20} & \scriptsize{5.09} & \scriptsize{4.24} & \scriptsize{4.87} & \scriptsize{\AA\ (FWHM)} \\
\hline
\rule[-1ex]{0pt}{2.0ex}  \scriptsize{Resolving Power $R$} & \scriptsize{1910} & \scriptsize{1140} & \scriptsize{1760} & \scriptsize{1920} & \scriptsize{($R = \lambda / \delta\lambda$)} \\
\hline
\rule[-1ex]{0pt}{2.0ex}  \scriptsize{Fringe Density} & \scriptsize{1770} & \scriptsize{776} & \scriptsize{923} & \scriptsize{797} & \scriptsize{lines mm$^{-1}$} \\
\hline
\rule[-1ex]{0pt}{2.0ex}  \scriptsize{Bragg Wavelength} & \scriptsize{400} & \scriptsize{590} & \scriptsize{750} & \scriptsize{940} & \scriptsize{nm} \\
\hline
\rule[-1ex]{0pt}{2.0ex}  \scriptsize{Angle of Incidence} & \scriptsize{12.04} & \scriptsize{6.57} & \scriptsize{11.17} & \scriptsize{12.28} & \scriptsize{\degree\ (on VPH Layer)} \\
\hline
\rule[-1ex]{0pt}{2.0ex}  \scriptsize{Fringe Tilt $\phi$} & \scriptsize{1.31} & \scriptsize{2.00} & \scriptsize{2.06} & \scriptsize{2.09} & \scriptsize{\degree\ (see Fig. \ref{fig:GrismSchem})} \\
\hline
\rule[-1ex]{0pt}{2.0ex}  \scriptsize{Assembly Physical Tilt $\theta_{\mathrm{grism}}$} & \scriptsize{15.3} & \scriptsize{14.3} & \scriptsize{15.3} & \scriptsize{15.3} & \scriptsize{\degree\ (see Fig. \ref{fig:GrismSchem})} \\
\hline
\rule[-1ex]{0pt}{2.0ex}  \scriptsize{Prism Wedge Angle $\gamma$} & \scriptsize{35.5} & \scriptsize{3.5} & \scriptsize{30.8} & \scriptsize{37.4} & \scriptsize{\degree\ (see Fig. \ref{fig:GrismSchem})} \\
\hline
\rule[-1ex]{0pt}{2.0ex}  \scriptsize{DCG Optical Thickness} & \scriptsize{3.5} & \scriptsize{6.0} & \scriptsize{4.9} & \scriptsize{7.5} & \scriptsize{$\mu$m} \\
\hline
\rule[-1ex]{0pt}{2.0ex}  \scriptsize{DCG Index Modulation} & \scriptsize{0.060} & \scriptsize{0.048} & \scriptsize{0.074} & \scriptsize{0.061} & \scriptsize{(assumed sinusoidal)} \\
\hline
\end{tabular}
\end{center}
\end{table}

%% file: t3.tex
\begin{table}[t!]
\caption{Minimum exposure times required for sky noise to equal read noise with LRS2 in dark sky conditions.} 
\label{tab:ExpTime}
\begin{center}       
\begin{tabular}{|r|c|c|c|c|c|c|c|c|} 
\hline
\rule[-1ex]{0pt}{3.0ex} \footnotesize{\textbf{Spectrograph Channel}} & \footnotesize{\textbf{UV}} & \footnotesize{\textbf{UV}} & \footnotesize{\textbf{Orange}} & \footnotesize{\textbf{Orange}} & \footnotesize{\textbf{Red}} & \footnotesize{\textbf{Red}} & \footnotesize{\textbf{Far Red}} & \footnotesize{\textbf{Far Red}} \\
\hline
\rule[-1ex]{0pt}{3.0ex} \footnotesize{\textbf{$\lambda$ (nm)}} & \footnotesize{375} & \footnotesize{420} & \footnotesize{500} & \footnotesize{610} & \footnotesize{690} & \footnotesize{790} & \footnotesize{870} & \footnotesize{970} \\
\hline
\rule[-1ex]{0pt}{3.5ex} \footnotesize{\textbf{Minimum Exposure Time (sec)}} & \footnotesize{2257} & \footnotesize{1082} & \footnotesize{602} & \footnotesize{351} &  \footnotesize{558} & \footnotesize{202} & \footnotesize{347} & \footnotesize{299} \\
\hline

\end{tabular}
\end{center}
\end{table}